\documentclass[12pt,preprint]{aastex}

\slugcomment{\it Astrophysical Journal, in press}
\shortauthors{Eisenhardt et al.}
\shorttitle{$z>1$ Galaxy Clusters}
 
\newcommand{\mum}{$\mu m$}
\newcommand{\muj}{$\mu$Jy}
\newcommand{\boot}{Bo\"otes}
 
\begin{document}
 
\title{Clusters of Galaxies in the First Half of the Universe\\
from the IRAC Shallow Survey}

\author{
Peter~R.~M.~Eisenhardt\altaffilmark{1},
Mark Brodwin\altaffilmark{2},
Anthony~H.~Gonzalez\altaffilmark{3},
S.~Adam Stanford\altaffilmark{4,5},
Daniel~Stern\altaffilmark{1},
Pauline Barmby\altaffilmark{6},
Michael~J.~I.~Brown\altaffilmark{7},
Kyle Dawson\altaffilmark{8},
Arjun Dey\altaffilmark{2},
Mamoru Doi\altaffilmark{9},
Audrey Galametz\altaffilmark{1,10},
B.~T.~Jannuzi\altaffilmark{2}, 
C.~S.~Kochanek\altaffilmark{11},
Joshua Meyers\altaffilmark{8,12}, 
Tomoki Morokuma\altaffilmark{9,13}, \&
Leonidas~A.~Moustakas\altaffilmark{1}}

\altaffiltext{1}{Jet Propulsion Laboratory, California Institute of
Technology, MS 169-327, 4800 Oak Grove Drive, Pasadena, CA 91109,
Peter.Eisenhardt@jpl.nasa.gov}

\altaffiltext{2}{National Optical Astronomy Observatory, 950 North Cherry Avenue, Tucson, AZ 85719}

\altaffiltext{3}{Department of Astronomy, University of Florida,
Gainesville, FL 32611}

\altaffiltext{4}{University of California, Davis, CA 95618}

\altaffiltext{5}{Institute of Geophysics and Planetary Physics,
Lawrence Livermore National Laboratory, Livermore, CA 94550}

\altaffiltext{6}{Department of Physics and Astronomy,
University of Western Ontario, 
1151 Richmond St., London, ON N6A 3K7}

\altaffiltext{7}{School of Physics, Monash University, Clayton, 
Victoria 3800, Australia}

\altaffiltext{8}{E.O. Lawrence Berkeley National
Laboratory, 1 Cyclotron Rd., Berkeley, CA 94720.}            

\altaffiltext{9}{Institute of Astronomy, Graduate
School of Science, University  of Tokyo 2-21-1 Osawa, Mitaka, Tokyo
181-0015, Japan}

\altaffiltext{10}{Observatoire Astronomique de Strasbourg, 
11 rue de l'Universit\'e, 67000 Strasbourg, France}

\altaffiltext{11}{Department of Astronomy, Ohio State University, 140 West 18th Avenue, 
Columbus, OH 43210}

\altaffiltext{12}{Department of Physics, University of California, Berkeley, CA 94720}

\altaffiltext{13}{National Astronomical Observatory of Japan,
2-21-1 Osawa, Mitaka, Tokyo 181-8588, Japan}

\begin{abstract}

We have identified 335 galaxy cluster and group candidates, 
106 of which are at $z > 1$,  using a 
$4.5\mu$m selected sample of objects from a 
7.25 deg$^2$ region in the 
{\it Spitzer} Infrared Array Camera (IRAC) Shallow Survey.  
Clusters were identified as 3-dimensional overdensities using 
a wavelet algorithm, based on 
photometric redshift probability distributions 
derived from IRAC and NOAO Deep Wide-Field Survey data. 
We estimate only $\sim10\%$ of the detections are spurious.
To date 12 of the $z > 1$ candidates have been confirmed spectroscopically, 
at redshifts from 1.06 to 1.41.  Velocity dispersions of
$\sim 750$ km s$^{-1}$ for two of these
argue for total cluster masses
well above $10^{14} M_\odot$, as does the mass estimated from
the rest frame near infrared stellar luminosity.  
Although not selected to contain a red sequence, some evidence
for red sequences is present in the spectroscopically confirmed
clusters, and brighter galaxies 
are systematically redder than the mean galaxy color 
in clusters at all redshifts.
The mean $I - [3.6]$ color for cluster galaxies up to $z \sim 1$ 
is well matched by a passively evolving model in which stars are
formed in a 0.1 Gyr burst starting at redshift $z_f = 3$.  
At $z > 1$, a wider range of formation histories is needed, 
but higher formation redshifts (i.e. $z_f > 3$) 
are favored for most clusters.

\end{abstract}


\section{Introduction\label{sec:intro}}

As the most massive gravitationally bound systems in the Universe,
the rate of emergence of galaxy clusters since the Big Bang might
be expected to be among the most straightforward predictions 
of cosmological models.  
Yet despite the advent of the era of precision cosmology
ushered in by observations of SNe and the cosmic microwave background
(CMB), significant uncertainty remains in the expected numbers of galaxy
clusters at $z > 1$.  The CMB temperature anisotropies on scales 
corresponding to clusters are not accurately known, leading to
a range of values for $\sigma_8$, the rms matter fluctuation in a sphere of
radius $8h^{-1}$ Mpc at $z=0$. Estimates of $\sigma_8$ vary significantly, 
including e.g. 
$0.67^{+0.18}_{-0.13}$ \citep{Gladders2007},
$0.76\pm0.05$ \citep{Spergel2007}, 
$0.80\pm0.1$ \citep{Hetterscheidt2007}, 
$0.85\pm0.06$ \citep{Hoekstra2006}, 
$0.90\pm0.1$ \citep{Spergel2003},
$0.92\pm0.03$ \citep{Hoekstra2002}, and 
$0.98\pm0.1$ \citep{BahcallBode2003}.
The range $\sigma_8 = 0.7$ to 1 
corresponds to a variation of a factor of nearly 20 
in the predicted numbers of $z > 1$ clusters
with $M_{\rm tot} > 10^{14} M_\odot$ \citep[e.g.][]{ShethTormen1999}.  
Removing this uncertainty
is a major goal of upcoming Sunyaev-Zeldovich cluster surveys such as the SZA 
\citep[Sunyaev-Zeldovich Array;][]{Loh2005},
AMI \citep[Arcminute Microkelvin Imager;][]{Kneissl2001}, 
ACT \citep[Atacama Cosmology Telescope;][]{Kosowsky2003}, 
and SPT \citep[South Pole Telescope;][]{Ruhl2004}.

\begin{figure}
\vspace*{-2cm}
\epsscale{1.15}
\plotone{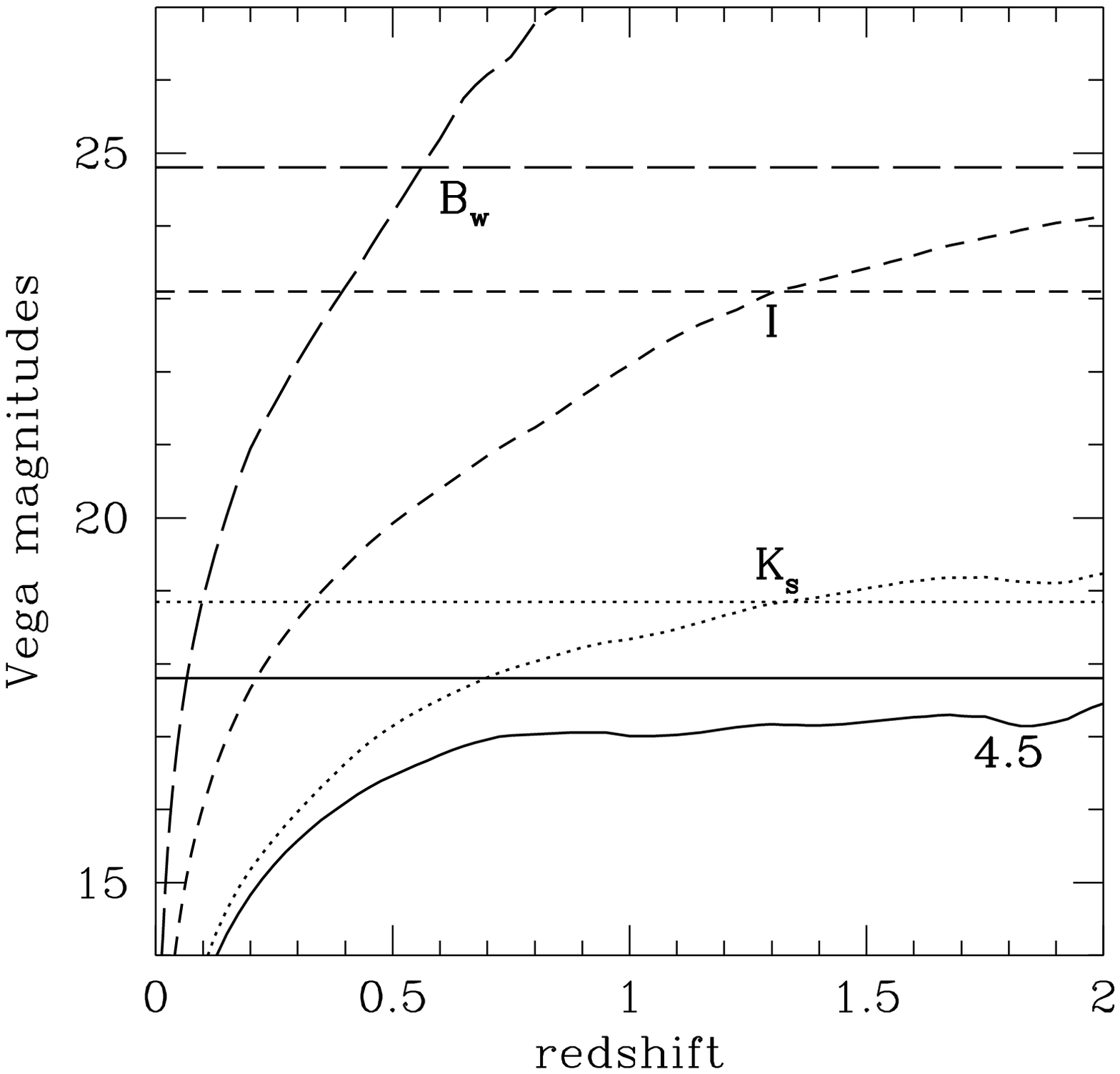}
\caption{
$L^*$ for cluster galaxies vs. $z$, in the observed $B_W$ (long dashed), 
$I$ (short dashed), $K_s$ (dotted), and IRAC $4.5\mu$m (solid) bands.  The
curves are based on a \citet{BC03} model where stars are formed 
in a 0.1 Gyr burst beginning at $z_f=3$ 
in a $\Lambda = 0.7$, $\Omega_m = 0.3$,  $h = 0.7$ cosmology,
which fits the observed $L^*$ in galaxy clusters to $z\ga 1$ 
\citep{DePropris1999}.  This model is referred to as the
``red spike'' model in the text.  
Horizontal lines show the $5 \sigma$ limits of the 
IRAC Shallow Survey, FLAMEX survey, and NDWFS 
in 5\arcsec\ diameter apertures.
Note a 5\arcsec\ aperture is larger
than optimum for detection at $B_W$ and $I$, and the 50\%
completeness limits in these bands are about 2 mag fainter.}
\label{m_vs_z}
\end{figure}

Since by definition\footnote{There is no generally 
agreed upon definition for galaxy 
clusters: see e.g. \citet{Abell1958}, \citet{Postman1996}, and
\citet{Rosati2002}. We define our criteria for candidate galaxy clusters
and groups in \S \ref{sec:detn}, and for spectroscopically confirmed
candidates in \S \ref{clusterz}.} 
galaxy clusters contain an unusually high density of galaxies,
they provide an efficient means of observing substantial numbers
of galaxies at a common distance, offering the hope of constructing the
analog of the Hertzsprung-Russell (i.e. color-magnitude) diagram for galaxy evolution.  
Indeed, studies of the relationship between color and magnitude 
indicate that clusters are the habitat of galaxies with the oldest
and most massive stellar populations 
(e.g.~\citealt*{SED98}; \citealt{Blakeslee2003}), 
objects reasonably free of the complications associated with
starbursts and dust.  These studies are consistent with an extremely
simple formation history for cluster galaxies, in which
their stars are formed in a short burst at high redshift, and
they evolve quiescently thereafter (we use the term 
``red spike model'' in referring to this scenario - see Figure \ref{m_vs_z}).  
Studies of the near-IR luminosity functions
of cluster galaxies reinforce this picture 
\citep[e.g.][]{DePropris1999, DePropris2007, Toft2004, Strazzullo2006}.  
With their large lookback times, therefore, high
redshift galaxy clusters also provide an observational pillar for our
understanding of the formation and evolution of galaxies.

Obtaining substantial samples  
of galaxy clusters at $z > 1$ has proved challenging, 
largely because such objects are difficult to detect using only optical data. 
Due to their greatly enhanced rate of star formation by $z \sim 1$, 
the UV emission from modest sized field galaxies 
overwhelms that from
the intrinsically red spectra 
of quiesecent, early type galaxies preferentially found in clusters.  
The Red Sequence Cluster Survey \citep{GladdersYee2000, GladdersYee2005}
uses the observed color-magnitude relationship in cluster galaxies to
improve the contrast and has proven highly efficient to $z \sim 1$, but the
optical colors of the red sequence become increasingly degenerate at
higher redshifts, as they no longer span the rest 4000\AA\ break.  
\citet{Wilson2006} describe a program to extend the red sequence technique
to higher redshift using Spitzer data, but it is also important to {\em test} for 
the existence of red sequences in $z > 1$ clusters rather 
than preselecting for them, if possible.

The contrast of high redshift clusters over the field improves at
longer wavelengths (Figure \ref{m_vs_z}), 
but the contrast against atmospheric
emission declines, and until recently the relatively small formats of
infrared detector arrays made surveying sufficient $z \ga 1$ volume a
formidable undertaking.  \citet{Stanford1997} reported the discovery of
a cluster at $z = 1.27$ in a 100 square arcminute survey to $K_s$(Vega)
= 20 (10$\sigma$).  But this survey required approximately 2 hours of
exposure in both $J$ and $K_s$ per position, and 30 allocated nights
of KPNO 4m time to complete. With estimates for the surface density
of $10^{14} M_\odot$ clusters at $z > 1$ in the range $0.2 - 4$
per square degree \citep{ShethTormen1999}, 
the discovery was in hindsight fortuitous.

Such considerations motivated a different approach, 
where extended sources in
deep X-ray surveys lacking prominent optical counterparts were targeted
for IR followup.  This technique yielded confirmed
clusters at $z = 1.10,$ 1.23, and 1.26 
\citep{Stanford2002, Rosati1999, Rosati2004}.
With the arrival of {\em XMM}, X-ray surveys offer renewed promise,
leading recently to the identification of galaxy clusters at $z = 1.39$
\citep{Mullis2005} and 1.45 \citep{Stanford2006}.  
With exposure times $> 20$ ksec and a 30 arcmin
field of view, a discovery rate of approximately 30 hours per candidate 
$z > 1$ cluster above $10^{14} M_\odot$ is expected (assuming one such
cluster per square degree, which corresponds to $\sigma_8 = 0.83$).

Searches for clusters around radio galaxies have yielded protoclusters
with redshifts as high as 4.1 
\citep{Pentericci2000, Venemans2002, Venemans2005} and possibly even 5.2 
\citep{Overzier2006}.
The very large redshifts of these systems
enable powerful inferences to be drawn regarding the formation of cluster
galaxies, but they are less useful as probes of the cosmological 
growth of structure. Lyman Break Galaxy
surveys with intensive followup spectroscopy on the Keck telescopes
have also identified highly overdense structures at $z=2.30$ and 3.09
\citep{Steidel1998, Steidel2005}, and \citet{Ouchi2007} discuss a
$z = 5.7$ structure identified via Ly-$\alpha$ emission in a narrow
band imaging survey.    

Recent advancements in IR detector array formats have renewed interest in
ground-based IR surveys.   In one example of the state of the art, 
Elston, Gonzalez et al. (2006) use the $2048 \times 2048$ pixel FLAMINGOS
camera to map 4 deg$^2$ to a 50\% completeness limit of
$K_s = 19.2$ (Vega). With 2 hour
exposures on the KPNO 2.1m each covering 1/10th deg$^2$, this leads
to an expected discovery rate for high redshift clusters per useful hour
of observing which is similar to {\em XMM}.  
The UKIDSS Ultra Deep Survey provides another recent example,
finding 13 cluster candidates with $0.6 < z < 1.4$
in a 0.5 deg$^2$ survey \citep{vanBreukelen2006}, one
of which has 4 spectroscopic redshifts at $z = 0.93$ 
\citep{Yamada2005}, and \citet{Zatloukal2007} find 12 candidates
with $1.23 < z < 1.55$ in a 0.66 deg$^2$ $H$-band survey in the COSMOS field. 
\citet{McCarthy2007} present a system with a high density of galaxies
with red optical to near-IR colors surrounding a galaxy at $z=1.51$,
identified in the 120 arcmin$^2$ Gemini Deep Deep Survey.  Candidates drawn 
from surveys of less than a square degree are unlikely to include many rich clusters, 
however.

With the launch of the {\it Spitzer Space Telescope} in 2003
\citep{Werner2004}, sensitive infrared arrays free from foreground thermal
emission were put into operation \citep{Fazio2004a}.  A major scientific
driver for the {\it Spitzer} Infrared Array Camera (IRAC) Shallow Survey
\citep{Eisenhardt2004} was the detection of $z > 1$ galaxy clusters.
The IRAC Shallow Survey uses 90 second exposures per position and covers
8.5 deg$^2$, leading to an expected discovery rate of $< 8$ hours per $z > 1$
cluster.  Here we present results from the IRAC Shallow Survey
cluster search, finding 106 cluster and group 
candidates at $z > 1$, of which we estimate only
$\sim10\%$ are spurious. 

A surface density of over 10 systems per square degree at $z > 1$ is
higher than expected for bound systems with masses above $10^{14} M_\odot$
for the current range of plausible $\sigma_8$ estimates.  
While we present evidence that at least two of the $z > 1$ clusters 
have masses well above $10^{14} M_\odot$, 
it is likely that our sample
includes systems with masses below $10^{14} M_\odot$ (i.e. groups), 
and perhaps some unbound filaments viewed end-on.  In the remainder
of this paper, for brevity the terms ``clusters" and ``candidates"  
are used to refer to all such objects which meet our selection criteria, 
unless otherwise stated.  

This paper describes how the cluster sample was identified, and
some of the overall photometric properties of the sample. 
We also provide new spectroscopic evidence supporting 
nine of these clusters, 
from $z = 1.057$ to 1.373. 
Spectroscopic evidence in support of IRAC Shallow Survey selected 
clusters at $z = 1.112,$ 1.243, and 1.413 was presented in \citet{Elston2006}, 
\citet{Brodwin2006}, and \citet{Stanford2005} respectively, and
we provide additional previously unpublished spectroscopy 
on those clusters here, for completeness.  
\citet{Brodwin2007} discuss the clustering of the clusters,
and Galametz et al. (in preparation) report on AGN incidence vs cluster-centric distance. 
Followup imaging with {\it HST} (GO 10496, Perlmutter; 10836, Stanford; and 
11002, Eisenhardt) and {\it Spitzer} (GO 30950, Eisenhardt) 
is underway, 
and future papers will examine the scatter in the color-magnitude 
relation as a function of morphological type, 
the dependence of cluster galaxy size on redshift,  
starburst activity in clusters vs. redshift,  
and the dependence of mean galaxy properties on surface density. 

A cosmology with $H_0 = 70$ km s$^{-1}$, $\Omega_m = 0.3$, 
$\Lambda = 0.7$ is assumed, and magnitudes are on the Vega system 
\citep[defined in][for IRAC]{Reach2005}.
At $z = 1 - 2$, this means that one arcminute corresponds to 
a physical scale of $480 - 502$ kpc, peaking at 508 kpc at $z = 1.6$.
Unless otherwise specified, physical (rather than co-moving) scales are
used throughout. 

\section{Data\label{sec:data}}

\subsection{IRAC Shallow Survey}

The IRAC Shallow Survey \citep{Eisenhardt2004} was designed to maximize
the number of reliable sources detected per unit time and to cover sufficient
area to detect significant numbers of $z \ga 1$ galaxy clusters.
As explained in \citet{Eisenhardt2004}, a 30 second exposure time per pointing is
close to optimum for maximizing source detections, and reliability was
obtained by requiring three independent exposures
separated by hours at each position.  The survey covers $\approx$ 8.5 deg$^2$ 
and reaches an aperture-corrected
5$\sigma$ depth of $\approx$ 19.1 and 18.3 mag (Vega) at 3.6 and $4.5 \mu$m in 
3\arcsec\ diameter apertures. It is a remarkable fact that 
90 seconds of combined exposure with IRAC 
on the 85 cm {\it Spitzer Space Telescope} provides sufficient
sensitivity to detect evolving $L^*$ galaxies to $z = 2$ (Figure 1).

\subsection{NOAO Deep Wide-Field Survey}

The survey was carried out in the Bo\"otes region of the
NOAO Deep Wide-Field Survey \citep[NDWFS;][]{JannuziDey1999}
to allow photometric redshifts to be derived using the deep optical 
imaging available for this field.  The NDWFS reaches 5$\sigma$
point-source depths in the $B_W, R,$ and $I$ bands of $\approx$ 27.1,
26.1, and 25.4 respectively.  
Typical exposure times per position with the Mosaic-1 camera on the
KPNO Mayall 4-m telescope were 1 -- 2 hours in $B_W$,
1 -- 2 hours in $R$, and 2 -- 4 hours in $I$, and the seeing ranged from
0.7 to 1.5\arcsec .  The data acquisition,
reduction, and catalog generation are discussed in detail by B. Jannuzi
et al. (in preparation) and A. Dey et al. (in preparation). This paper
uses the NDWFS third data release (DR3) images and SExtractor catalogs 
which can be obtained
through the NOAO data archive\footnote{http://www.noao.edu/noao/noaodeep}.

\subsection{AGN and Galaxy Evolution Survey, FLAMINGOS Extragalactic Survey, and Other Surveys}

The AGN and Galaxy Evolution Survey (AGES, C. Kochanek et al. in preparation)
provides spectroscopic redshifts for $\approx 17,000$ objects 
(using the version 2.0 catalog) in the IRAC
Shallow Survey.  AGES is highly complete for sources 
brighter than 15.7 mag at $4.5 \mu$m, 
and also for sources brighter than $I=18.5$, with many redshifts
for sources up to $I=20$,
enabling excellent assessment of the photometric redshifts
to $z \sim 0.5$ (Figure 1). 

Deep near infrared imaging
from the FLAMINGOS Extragalactic Survey (FLAMEX)
is available for half the Bo\"otes region \citep{Elston2006}, and
was used for deriving a prior on the redshift likelihood functions
(see \S \ref{sec:photz}).

Imaging of the Bo\"otes NDWFS
field has also been obtained 
in the radio \citep{deVries2002},
at 24, 70, and $160 \mu$m with the MIPS instrument on {\it Spitzer} 
\citep{Houck2005}, in the $z-$band \citep{Cool2007}, 
in the $UV$ with $GALEX$, and  
in X-rays to a depth of 5 ksec with the ACIS instrument on the 
{\it Chandra X-ray Observatory} \citep{Murray2005}, 
but these data are not used in this paper.

\section{Sample\label{sec:sample}}

Object selection for the cluster search was carried out in the
$4.5\mu$m band, because the negative K-correction as the rest--frame
$1.6\mu$m peak shifts into this band leads to a flux which is
nearly independent of redshift for $0.7 < z < 2$ (Figure 1).  
Object detection and photometry was carried out
using SExtractor \citep{sextractor} in double--image mode, allowing
matched aperture photometry in the other IRAC bands. 
While smaller apertures maximize the depth of the survey, 
Monte Carlo simulations showed that
5\arcsec\ diameter aperture magnitudes are necessary to provide 
sufficiently reliable color measurements and photometric redshifts 
\citep{Brodwin2006}.   
This flux limit (5$\sigma$ at $4.5 \mu$m in 5\arcsec)
corresponds to $13.3$ \muj, or a
Vega--based magnitude of $17.8$.

\subsection{Matching to the NDWFS Catalog}

The NDWFS catalogs were also generated using SExtractor, but run in
single-image mode in each band. Detections in the different optical
bands and between the optical and IRAC catalogs were matched if the
centroids were within $1\arcsec$ of each other, using the closest optical
source if more than one satisfied this criterion. For very extended objects
(generally at $z\le 0.2$),
detections in the different bands were matched if the centroids were
within an ellipse defined using the second order moments of the light
distribution of the object \citep{Brown2005}.

\subsection{$\mathbf B_W$RI[3.6] Flux Limits and 
Photometric Errors\label{sec:phot-limits}}

The NDWFS data were taken over
several years in variable conditions, and therefore the photometric
depths vary somewhat from pointing to pointing.  Average 50\%
completeness limits for the $B_W$, $R$, and $I$--bands were
26.7, 25.6, and 25.0 respectively.  Average 5$\sigma$ flux
(magnitude) limits in a 5\arcsec aperture (which is significantly larger
than the optimum detection aperture for these data) were measured via Monte
Carlo simulations to be $0.45$ \muj\ ($24.8$ mag) in $B_W$,
$1.03$ \muj\ ($23.7$ mag) in $R$, and $1.45$ \muj\ ($23.1$ mag) in the
$I$--band \citep{Brodwin2006}.  
In the space--based IRAC Shallow survey the depth is more
uniform, and in the $3.6\mu$m band it is $10.0$ \muj\ ($18.6$
mag), also derived from a Monte Carlo simulation.

Although the majority of objects in the $4.5\mu$m-selected catalog are
well detected at shorter wavelengths, this is not generally the
case for $z>1$ red ellipticals.  Objects at these redshifts 
are often quite faint in the optical as the 4000\AA\ break is longward of
the $I$--band (see Figure 1).  
Where sources were observed but not detected, the flux
was taken to be zero and a Monte Carlo 1$\sigma$ error was adopted.
This approach is optimal for photometric redshift fitting, where the
non-detection provides important contraints on the galaxy spectral
energy distribution (SED).

\subsection{Sample for Photometric Redshift Estimation 
and Cluster Search\label{sec:cluster_search_sample}}

The photometry
used for photometric redshift estimation (\S \ref{sec:photz})
consists of $B_WRI[3.6][4.5]$ data
with the Monte Carlo
photometric errors and limits noted above.
The $5.8 \mu$m and $8.0 \mu$m bands were not used  
because they do not have the sensitivity to
detect $z>1$ cluster $L^*$ galaxies in the IRAC Shallow Survey. 
While the IRAC Shallow Survey covers 8.5 $\deg^2$ in each band,
the overlap area observed in both the 3.6 and $4.5 \mu$m IRAC bands 
is 8.0 $\deg^2$.  All of this falls within the optical NDWFS \boot~region. 
However, due to haloes around bright objects, shorter observation
times for regions at the edges of individual mosaic camera pointings, 
and residual CCD defects, 0.75 $\deg^2$
may have lower quality optical photometry and hence photometric redshifts. 
Hence these regions are also excluded from the sample. 

Finally, 14,044 stars were removed from the catalog 
using the SExtractor stellarity index in the
best seeing optical data for each location.   Comparison to the
star count model of \citet{Arendt1998}, as tabulated 
in \citet{Fazio2004b}, suggests that $\sim80\%$ of stars 
were identified via this approach, 
leaving $\sim 2\%$ of the $4.5\mu$m sample as unrecognized stars.  

In summary, photometric redshifts were estimated
for a total of 175,431 objects 
brighter than $13.3$ \muj\ ($5\sigma$) at $4.5 \mu$m in 5\arcsec\  
in a 7.25 $\deg^2$ region.

\section{Photometric Redshifts\label{sec:photz}}

A full description of the photometric redshift methodology is
given in \citet{Brodwin2006}. 
A summary of those aspects most relevant to cluster detection 
is provided here. 
Photometric redshifts were computed using an empirical
template--fitting algorithm which linearly interpolates between   
the four Coleman, Wu, \& Weedman (1980) SEDs (E, Sbc, Scd, and Im), 
augmented by the \citet{Kinney1996} SB3 and SB2 starburst templates.  
These SEDs were extended to the far--UV and near--IR
using \citet{BC03} models. These stellar photospheric models 
do not include emission from dust,
in particular 
the PAH features which dominate the $3 < \lambda < 12 \mu$m portion of
the spectrum in starforming galaxies, but at $z > 1$ the photometry
used does not sample these rest wavelengths. 

In addition to the large AGES (C. Kochanek et al. in preparation) survey, 
there are $\sim 500$ spectroscopic redshifts
extending to $z\sim1.5$ gleaned from several ongoing surveys in the
Bo\"otes field.  These were used as training sets to adjust the templates
and photometric zero points to improve overall redshift accuracy and
reliability \citep[see][for details]{Brodwin2006}.   
Comparison with these spectroscopic samples
shows that an rms dispersion of $\sigma_z \sim 0.06(1+z)$ is
achieved for 95\% of galaxies to at least $z=1.5$.  
Subsequent follow--up spectroscopy of high redshift candidate clusters 
\citep[\S \ref{sec:Keck}; see also ][]{Stanford2005, Brodwin2006, Elston2006} 
verify this accuracy.

A key output of the \citet{Brodwin2006} technique is the redshift
probability function for each object, $P(z)$, derived directly from
the redshift-axis projection of the full redshift--SED likelihood surface.  
To transform these simple redshift likelihood functions into true 
probability distribution functions, a prior consisting of the observed
redshift distribution was applied.  This was measured in the FLAMEX 
region using high--quality $B_WRIJK_s[3.6][4.5]$ photometric redshifts.  
Comparison to the spectroscopic sample illustrates that the resulting
distribution functions are statistically valid 
in the sense that integrated areas
accurately represent redshift probabilities at the 1, 2, and 3$\sigma$ 
levels.  These $P(z)$ functions
are the input to the wavelet detection algorithm discussed in \S 
\ref{sec:detn} and are also used in \S \ref{sec:discussion}.

Note that the $P(z)$ distributions shows relatively little dependence
on galaxy type \citep{Brodwin2006}.  
The reason is that while all galaxies are selected to have 
at least 5 sigma detections in [4.5], red galaxies are often 
only marginally detected or even undetected in our optical images, since 
they have very little blue light.  This is particularly true at $z>1$ 
where the $k$-correction for early types is large in the optical. 
Blue galaxies have good $B_wRI$[3.6][4.5] photometry, because they 
have more blue light.  Thus even though blue, late-type galaxy SED's 
have smaller breaks, they have more extensive (useful) photometry.  
These compensating effects lead to effectively type-independent 
photometric redshifts for $z>1$ galaxies.  

\section{Cluster Detection\label{sec:detn}}

We employed a wavelet analysis to identify galaxy clusters within the
\boot~region.  Wavelet decomposition is a commonly used technique
for cluster identification in X-ray images 
\citep[for example, see][]{Vikhlinin1998,Valtchanov2004,Andreon2005,Kenter2005}, 
where it provides an effective means of identifying extended sources 
in the presence of contaminating point sources. 
In principle, a similar analysis can be used 
with optical and infrared data sets, using
galaxies rather than X-ray photons to identify extended sources. 
As is well known, galaxy number counts are more susceptible to projection
effects than is bremsstrahlung emission from the ICM. 
This issue is one that must be dealt with 
for all optical and infrared cluster searches, 
and consequently most such searches for distant clusters 
make explicit assumptions about the
properties of the distant cluster population, 
such as an assumed density profile
\citep[e.g.,][]{Postman1996,Olsen1999,Scodeggio1999}  
or the presence of a red sequence \citep{GladdersYee2000,GladdersYee2005}. 
The SDSS C4 technique \citep{Miller2005} does not require a {\em red}
sequence, but it does demand that the colors of cluster
galaxies are similar to one another.

A significant advantage of the \boot~data set 
is that the photometric redshifts
permit such assumptions to be minimized.  
The full photometric redshift probability distributions
$P(z)$ were used 
to construct weighted galaxy density maps within
overlapping redshift slices of width $\Delta z=0.2$, 
stepping through redshift space in increments of $\delta z=0.1$.  
For each galaxy the weight in the map corresponds 
to the probability that the galaxy lies within the given redshift slice. 
Weighting in this fashion de-emphasizes sources 
for which the redshift is poorly constrained.
It is worth emphasizing that {\it all} galaxies 
were included in construction of the density map, 
and consequently cluster detection should be 
relatively independent of SED type and hence independent of 
morphology. 
Finally, cluster detection will be insensitive to 
the resolution of the density maps provided that the
pixel size is small compared to the angular extent 
of cluster cores at all redshifts. 
A resolution of 12$\arcsec\ (\sim 100$ kpc) per pixel was used, 
which satisfies this criterion while being 
sufficiently large to keep computational overhead manageable.

Galaxy cluster candidates were detected within each redshift slice 
by convolving the density map with the wavelet kernel. 
We use a Gaussian difference kernel of the form
\begin{equation}
k(r) = \frac{e^{-r^2/(2 \sigma_1^2)} }{ \sigma_1^2}  - \frac{e^{-r^2/(2 \sigma_2^2)} }{ \sigma_2^2},
\end{equation}
where $\sigma_1=400$ kpc and $\sigma_2=1600$ kpc, and which crosses zero near
$r = 1$ Mpc. The scale of
the kernel is fixed in physical rather than angular units, preserving
our ability to uniformly identify comparable systems at different redshifts.
The precise physical values of $\sigma$ are subject to refinement, but
the selected values effectively isolate overdensities on the scale of clusters or groups. 

Galaxy cluster candidates were detected in each redshift slice of 
these wavelet smoothed galaxy density maps using a simple  
peak-finding algorithm. To establish a consistent significance level 
for the candidates, 1000 bootstrap simulations were carried out within 
each redshift slice. The existing P(z) distributions, right 
ascensions, and declinations were repeatedly shuffled, convolved with the 
wavelet kernel, and candidates detected to find the threshold 
corresponding to one false positive per redshift slice within the 
\boot~field. A list of detections above this significance threshold was generated 
for each redshift slice.  Because the right ascensions and declinations were 
shuffled independently, the true correlated background was not preserved. 
Consequently, the contamination rate may be somewhat higher 
than the one false positive per redshift slice expected if we had 
preserved the correlated background. For the current analysis 
we accept the somewhat higher contamination rate in exchange 
for improved completeness, particularly for the highest redshift clusters.  
The majority of clusters are detected 
in multiple redshift slices (a natural consequence of sampling 
in slices separated by step sizes finer than the galaxy redshift 
uncertainties). Multiple detections with small separations in 
positions and redshift slices were considered to be a single cluster, 
which was assigned an estimated redshift and position corresponding 
to the slice with the highest statistical significance. 
A total of 335 candidates were found in redshift slices 
with centers from 0.1 to 1.9, including 98 in slices with $z>1$.  

While the detection technique is on three dimensional overdensities, 
it is not immune to projection effects.  
The expected number of clusters in each cylindrical bin of
length $\Delta z=0.2$ and radius 1 Mpc (the radius of the wavelet
detection kernel) was computed to quantify the extent of projection.  
This calculation included both the random
expectation due to the observed number density and the excess clusters
expected due to the observed clustering of this sample \citep{Brodwin2007}.  
The projection rate is most significant
for lower redshifts because of the large angular size corresponding
to 1 Mpc, affecting $\sim 10\%$ of systems at $z=0.5$ to $\sim 20\%$ 
at $z = 0.1$.  The projection rate decreases to below $\sim
4$\% at $z =1$ and is negligible at $z > 1.5$.  Overall we estimate
that $\sim10\%$ of the candidates may be spurious, including the
contamination noted in the previous paragraph.

\section{Cluster Redshifts and Members\label{clusterz}}

The peak of the summed $P(z)$ distribution at the cluster detection
location was taken as the initial estimate of each cluster's redshift. 
To improve this estimate, individual objects within a 1 Mpc radius of
the cluster which included the cluster redshift within the $1\sigma$
range of their $P(z)$ functions were considered candidate cluster
members.  A refined estimate ($z_{\rm est}$) of the mean cluster
redshift was obtained from the peak of the summed $P(z)$ distribution
for these members.  From this analysis, 104 candidates have $z_{\rm est} > 1$.   

In Table \ref{z1table} and Figures 2 - 13 we present 12 of the cluster
candidates that were detected via the above criteria, and subsequently
spectroscopically confirmed at the Keck Observatory to lie at $z>1$
(see \S \ref{sec:Keck}).  
Two of these, ISCS~J1434.1+3328 and ISCS~J1429.2+3357, have $z_{\rm est} = 0.98$ 
but the spectroscopic mean redshifts are $z = 1.057$ and $1.058$ respectively. 
Hence we report a total of 106 clusters at $z > 1$, of which 
roughly 10\% may be expected to arise by chance 
or from projection effects, for the reasons noted in \S \ref{sec:detn}.
Column 1 of Table 1 provides
the catalog number of each cluster. The catalog numbers 
increase with decreasing detection
significance (\S \ref{sec:detn})\footnote{These numbers extend beyond 335 
for continuity with an earlier, preliminary version of the
catalog used to plan spectroscopic and other followup observations.}.  
Column 2 is the IAU designation for each cluster, based on the
(J2000) coordinates of the detection given in columns 3 and 4.  
Column 5 provides the $z_{\rm est}$ value described above, and 
column 6 gives the mean redshift of the spectroscopically confirmed members.  
Column 7 provides the  
number of photometric redshift members of the cluster,
defined as galaxies within 1 Mpc of the cluster center and
with integrated $P(z) \ge 0.3$ in the range $z_{\rm est} \pm 0.06(1 + z_{\rm est})$. 
These galaxies are used to calculate mean cluster colors in \S \ref{sec:Im3p6_vs_z}. 
Column 8 reduces the number in column 7 by the number of galaxies which satisfy these criteria for each cluster redshift over the entire field,
scaled to the 1 Mpc radius area for each cluster redshift. 
Column 9 gives the number of spectroscopically confirmed member galaxies in each cluster (see below). 
Column 10 gives the mean $I - [3.6]$ color for the photometric redshift member galaxies, and 
column 11 provides the sum of their luminosities in [4.5] 
relative to an $L^*$ value which evolves according to the ``red spike'' model 
shown in Figure \ref{m_vs_z}, corrected for the average over the field.
Column 12 gives the luminosity in [4.5] of the brightest photometric redshift 
member galaxy relative to
$L^*$. 

We define a $z > 1$ cluster as spectroscopically confirmed if it contains
at least 5 galaxies in the range $z_{\rm est} \pm 0.06(1 + z_{\rm est})$ 
and within a radius of 2 Mpc, 
whose spectroscopic redshifts match to within $\pm 2000(1 + \left<z_{\rm sp}\right>)$ km/s.
The spectroscopic redshifts must also be of class A or B. 
Class A spectra have unambiguous redshift determinations, 
typically relying upon multiple well detected emission or absorption features.  
Class B spectral features are reliable but are less well detected.  
The radius threshold for spectroscopic membership is larger than the 1 Mpc used for
photometric redshift members for practical reasons:  
most of the $z > 1$ redshifts reported here were obtained with slitmasks 
extending out to approximately 2 Mpc from the cluster center.  Typically $< 10$
photometric redshift members within 1 Mpc could be accommodated in the
mask design, and often no redshift could be determined from the resulting spectra.

The least significant cluster in Table 1 is 
ISCS~J1434.5+3427, which is the 327th most significant detection
out of the sample of 335, and the 100th most significant detection at
$z > 1$.  Despite its relatively low detection significance,
ISCS~J1434.5+3427 has a striking filamentary morphology (Figure
\ref{color10.342}), and has eleven 
spectroscopically confirmed members \citep[see also][]{Brodwin2006}.

\clearpage

\thispagestyle{empty}

\begin{deluxetable}{lcrccccccccc}
\tabletypesize{\normalsize}
\tablecaption{$z > 1$ ISCS Clusters with Spectroscopic Confirmation\tablenotemark{a} 
\label{z1table}}
\tablewidth{0pt}
\tablehead{
\colhead{Rank} & 
\colhead{ID} & \colhead{RA} & \colhead{Dec} 
& \colhead{$z_{\rm est}$}
& \colhead{$\left<z_{\rm sp}\right>$} & \colhead{${\rm N}_{\rm p}$} & 
\colhead{${\rm N}_{\rm p}(\rm bc)$} & \colhead{${\rm N}_{\rm sp}$} &
\colhead{$\left<I - [3.6]\right>$} &
\colhead{$L_{\rm tot}/L^*$} & \colhead{$L_{\rm bcg}/L^*$} 
}
\rotate
\startdata

152 & ISCS\_J1434.1+3328 & 14:34:10.37 & +33:28:18.3 & 0.98 & 1.057 & 32 & 13 &  6 & 4.83 & 15 & 2.5 \\
51  & ISCS\_J1429.2+3357 & 14:29:15.16 & +33:57:08.5 & 0.98 & 1.058 & 45 & 26 &  7 & 4.80 & 34 & 4.2 \\
19  & ISCS\_J1433.1+3334 & 14:33:06.81 & +33:34:14.2 & 1.02 & 1.070 & 57 & 38 & 20 & 4.93 & 45 & 4.3 \\
123 & ISCS\_J1433.2+3324 & 14:33:16.01 & +33:24:37.4 & 1.01 & 1.096 & 31 & 11 &  6 & 4.94 & 15 & 4.6 \\
17  & ISCS\_J1432.4+3332 & 14:32:29.18 & +33:32:36.0 & 1.08 & 1.112 & 49 & 31 & 23 & 5.16 & 47 & 5.3 \\
34  & ISCS\_J1426.1+3403 & 14:26:09.51 & +34:03:41.1 & 1.08 & 1.135 & 31 & 13 &  7 & 5.18 & 26 & 4.0 \\
14  & ISCS\_J1426.5+3339 & 14:26:30.42 & +33:39:33.2 & 1.11 & 1.161 & 52 & 35 &  5 & 5.15 & 47 & 2.8 \\
342 & ISCS\_J1434.5+3427 & 14:34:30.44 & +34:27:12.3 & 1.20 & 1.243 & 27 & 12 & 11 & 5.51 & 27 & 6.0 \\
30  & ISCS\_J1429.3+3437 & 14:29:18.51 & +34:37:25.8 & 1.14 & 1.258 & 23 &  7 &  9 & 5.30 & 11 & 6.5 \\
29  & ISCS\_J1432.6+3436 & 14:32:38.38 & +34:36:49.0 & 1.24 & 1.347 & 30 & 17 &  8 & 5.65 & 31 & 4.3 \\
25  & ISCS\_J1434.7+3519 & 14:34:46.33 & +35:19:33.5 & 1.37 & 1.373 & 19 &  9 &  5 & 5.77 & 17 & 3.9 \\
22  & ISCS\_J1438.1+3414 & 14:38:08.71 & +34:14:19.2 & 1.33 & 1.413 & 25 & 15 & 10 & 5.75 & 24 & 3.5 \\
\enddata
\tablenotetext{a}{Columns are explained in \S \ref{clusterz}}
\end{deluxetable}

\clearpage

\begin{figure}[bthp]
\epsscale{1.0}
\plotone{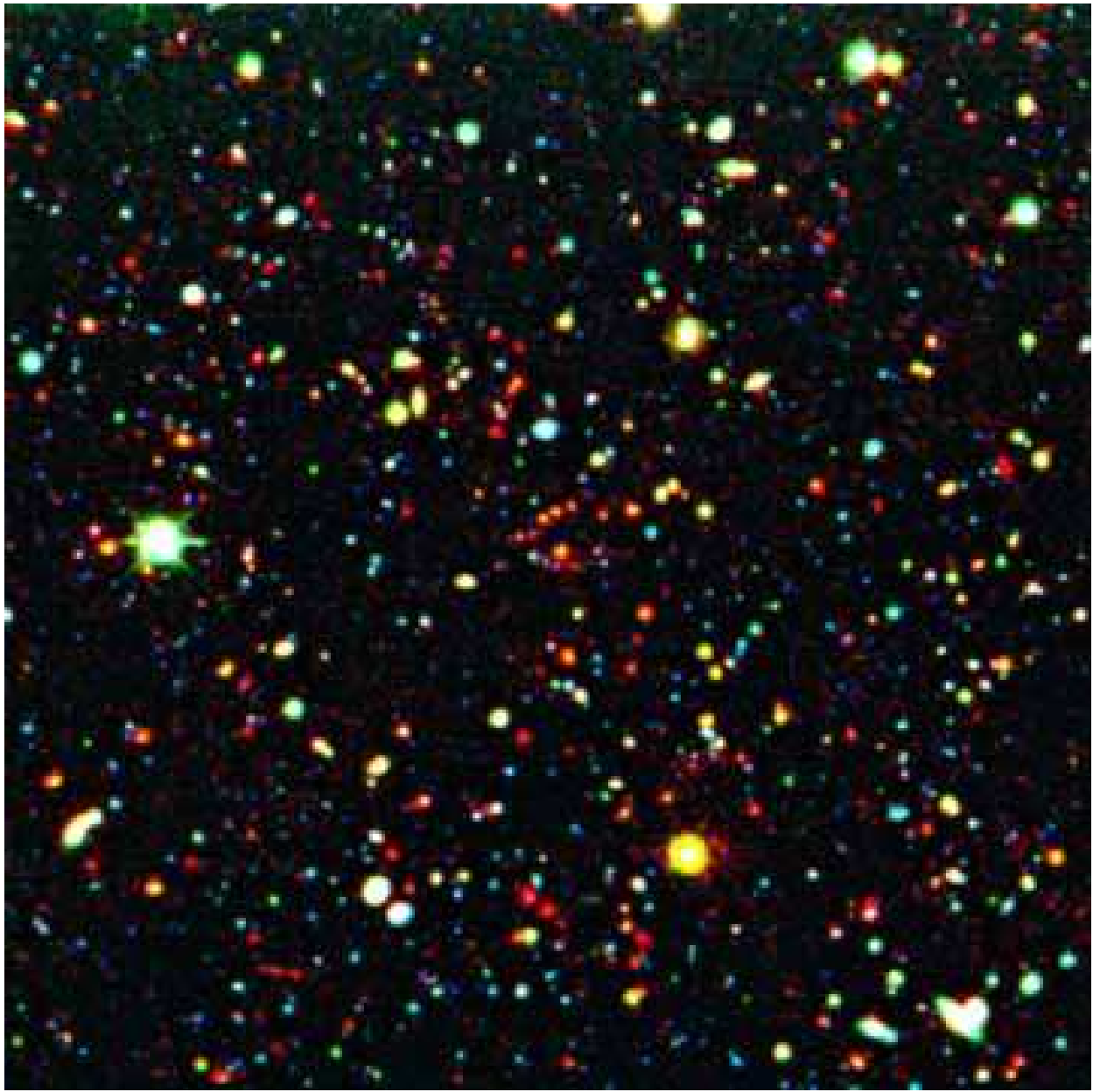}
\caption{Composite NDWFS and {\it Spitzer} IRAC 
$B_W$, $I$, $[4.5]$ color image of 
cluster ISCS~J1434.1+3328 at $\left<z_{\rm sp}\right>=1.057$.
North is up and east is left, and the field size is 5\arcmin\ square 
($\sim 2.4 - 2.5$ Mpc for $z = 1 - 1.5$).  The published version of these
figures will label photometric redshift members, and spectroscopic members
and non-members.}
\label{color10.152}
\end{figure}

\begin{figure}[bthp]
\plotone{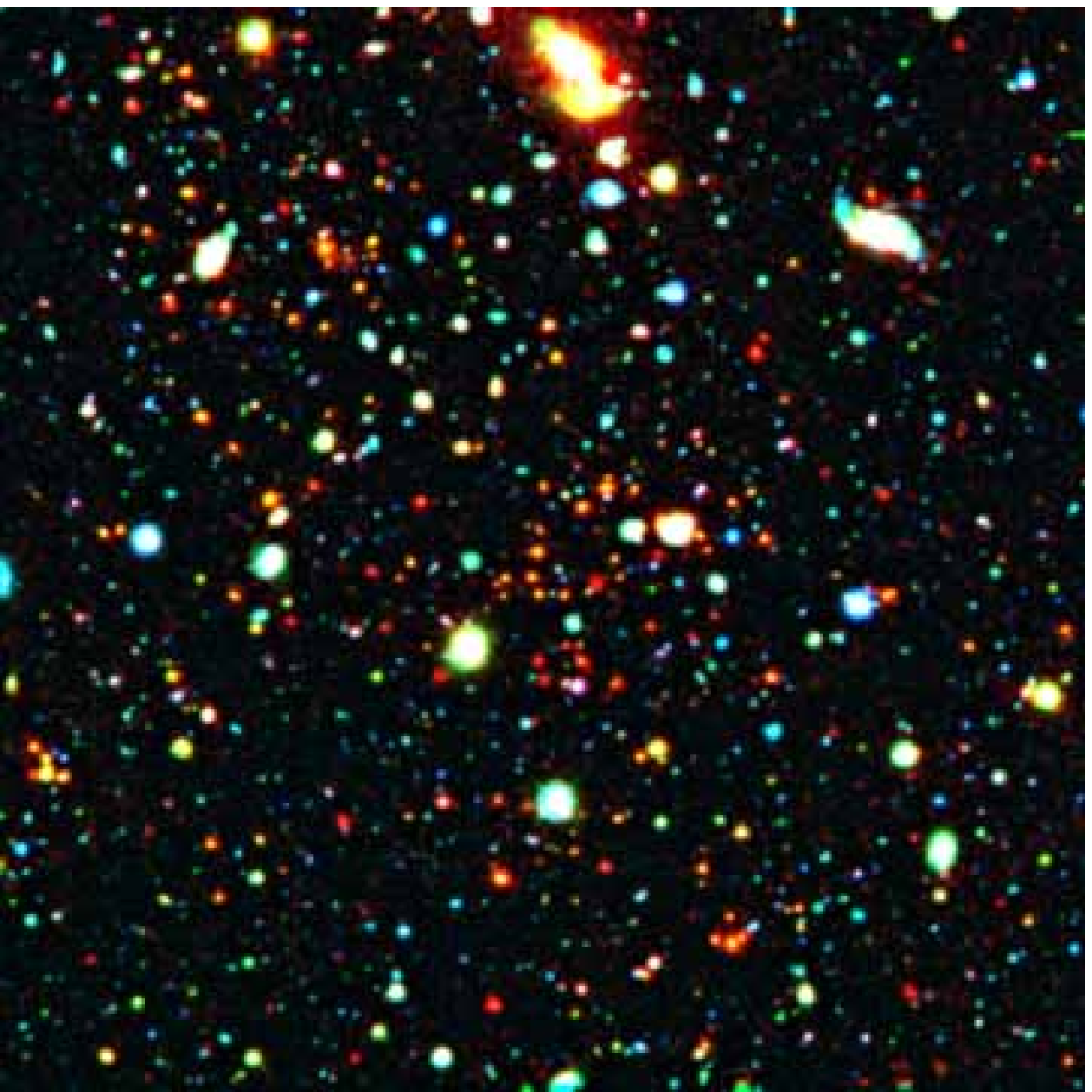}
\caption{
As for Figure \ref{color10.152}, but for 
cluster ISCS J1429.2+3357 at $\left<z_{\rm sp}\right>=1.058$.}
\label{color10.51}
\end{figure}

\begin{figure}[bthp]
\plotone{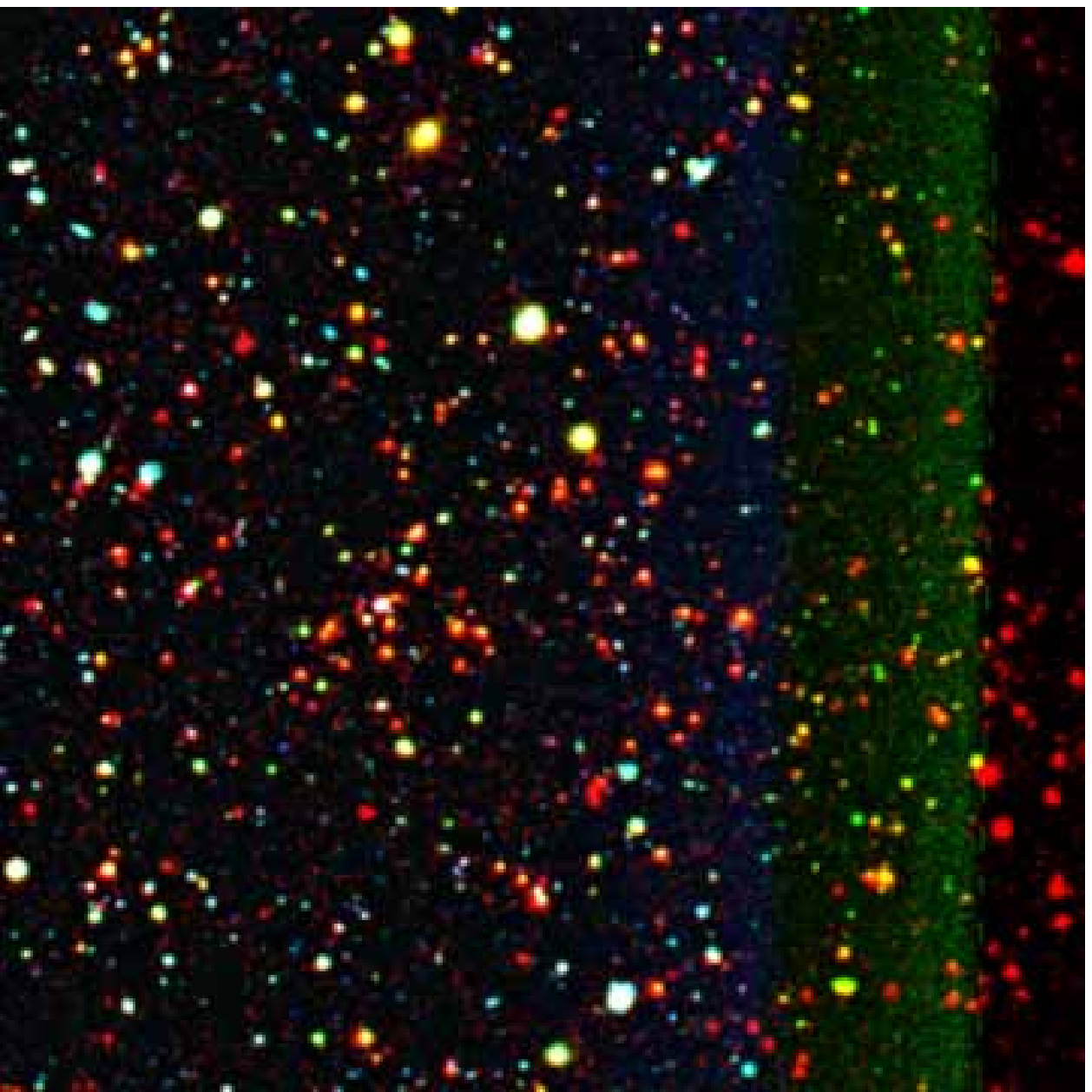}
\caption{
As for Figure \ref{color10.152}, but for 
cluster ISCS J1433.1+3334 at $\left<z_{\rm sp}\right>=1.070$.}
\label{color10.19}
\end{figure}

\begin{figure}[bthp]
\plotone{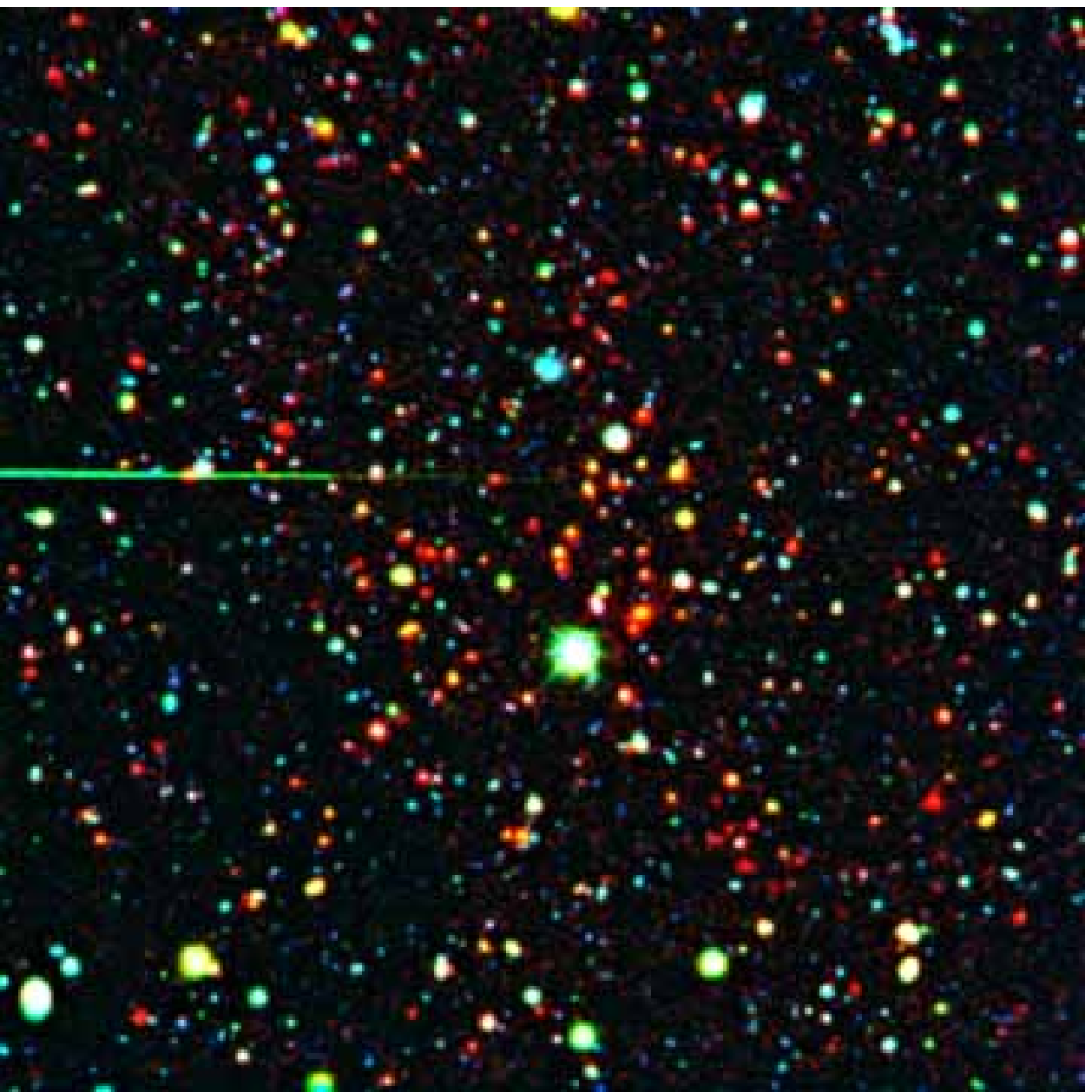}
\caption{
As for Figure \ref{color10.152}, but for 
cluster ISCS J1433.2+3324 at $\left<z_{\rm sp}\right>=1.096$.}
\label{color10.123}
\end{figure}

\begin{figure}[bthp]
\plotone{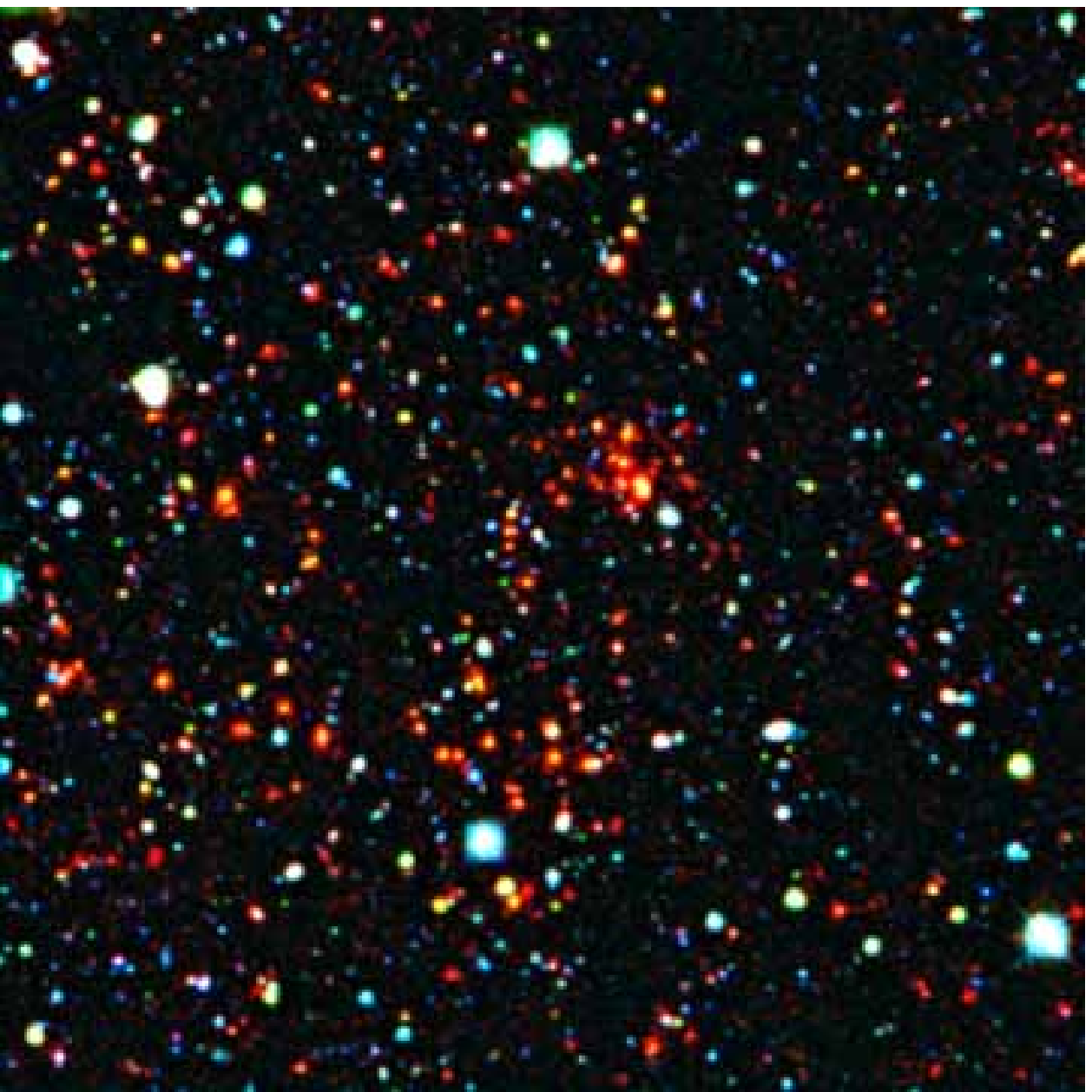}
\caption{
As for Figure \ref{color10.152}, but for 
cluster ISCS J1432.4+3332 at $\left<z_{\rm sp}\right>=1.112$.}
\label{color10.17}
\end{figure}

\begin{figure}[bthp]
\plotone{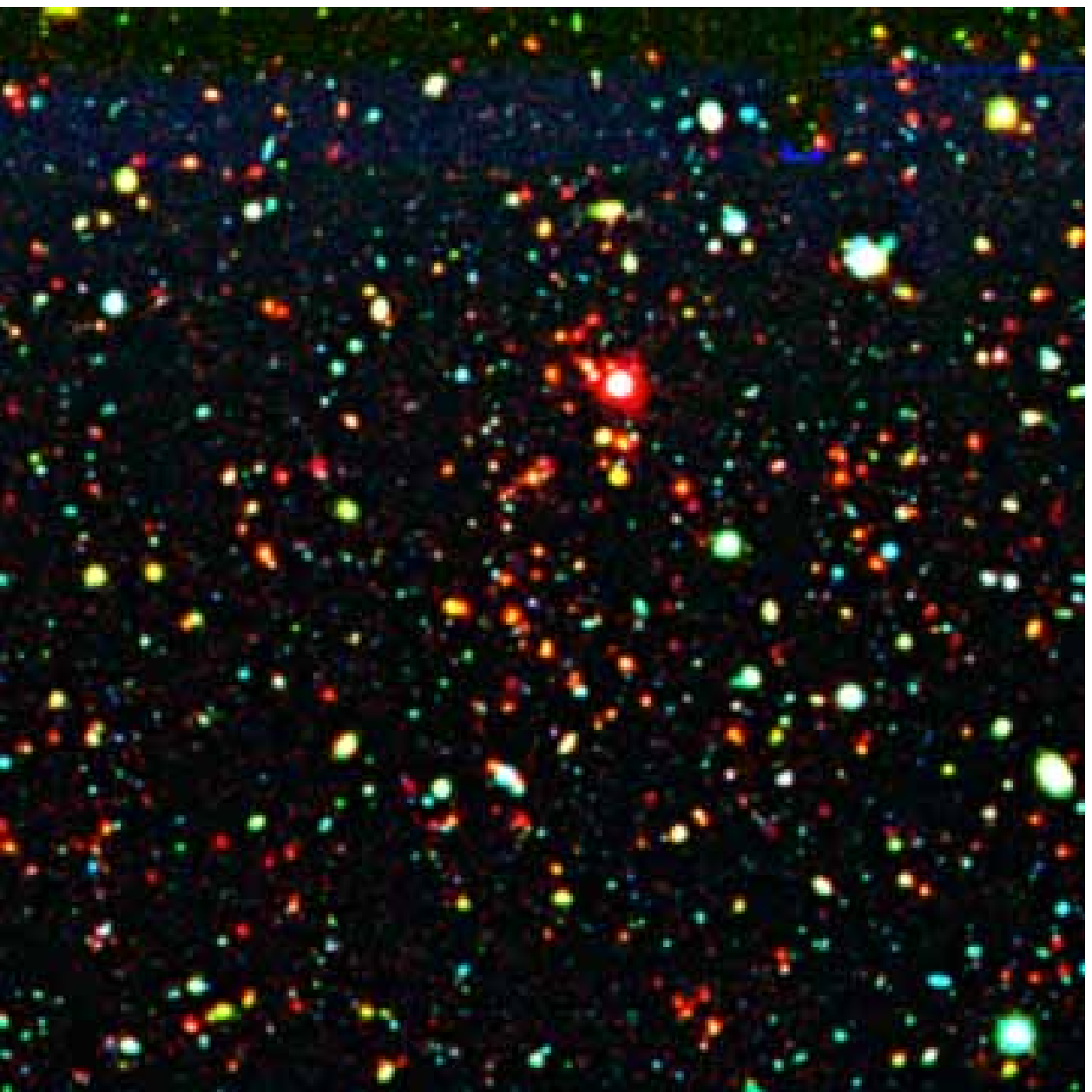}
\caption{
As for Figure \ref{color10.152}, but for 
cluster ISCS J1426.1+3403 at $\left<z_{\rm sp}\right>=1.135$.}
\label{color10.34}
\end{figure}

\begin{figure}[bthp]
\plotone{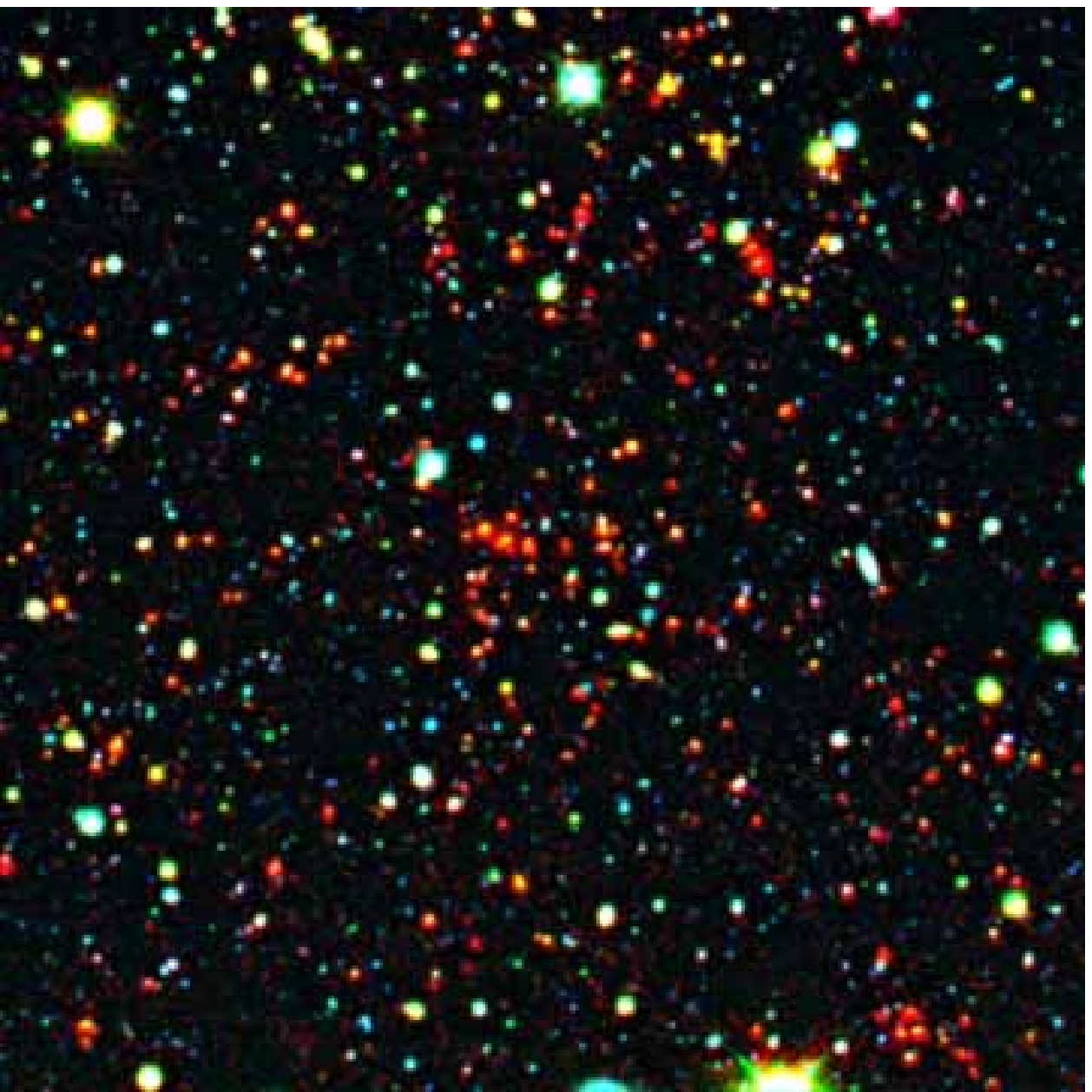}
\caption{
As for Figure \ref{color10.152}, but for 
cluster ISCS J1426.5+3339 at $\left<z_{\rm sp}\right>=1.161$.}
\label{color10.14}
\end{figure}

\begin{figure}[bthp]
\plotone{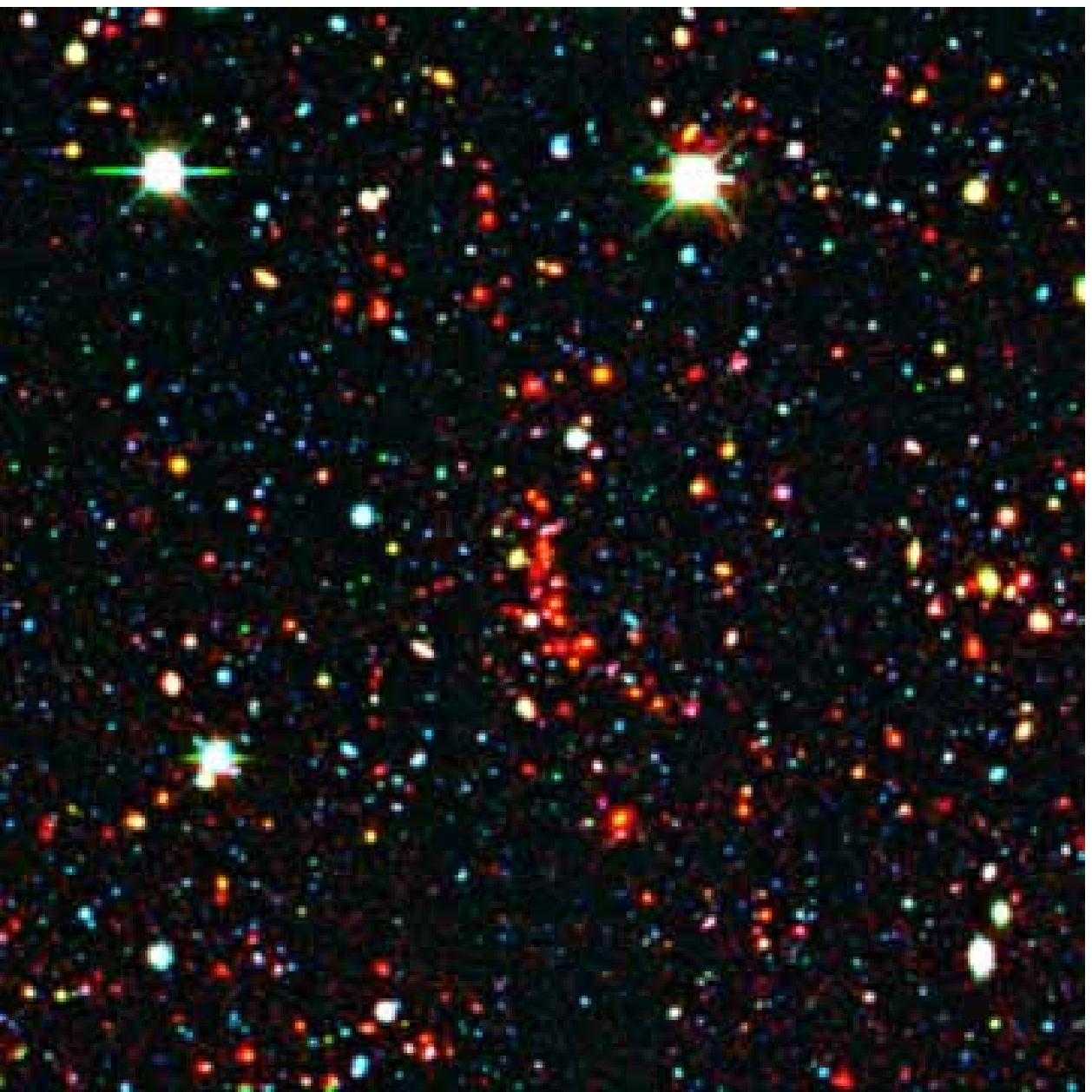}
\caption{
As for Figure \ref{color10.152}, but for 
cluster ISCS J1434.5+3427 at $\left<z_{\rm sp}\right>=1.243$.}
\label{color10.342}
\end{figure}

\begin{figure}[bthp]
\plotone{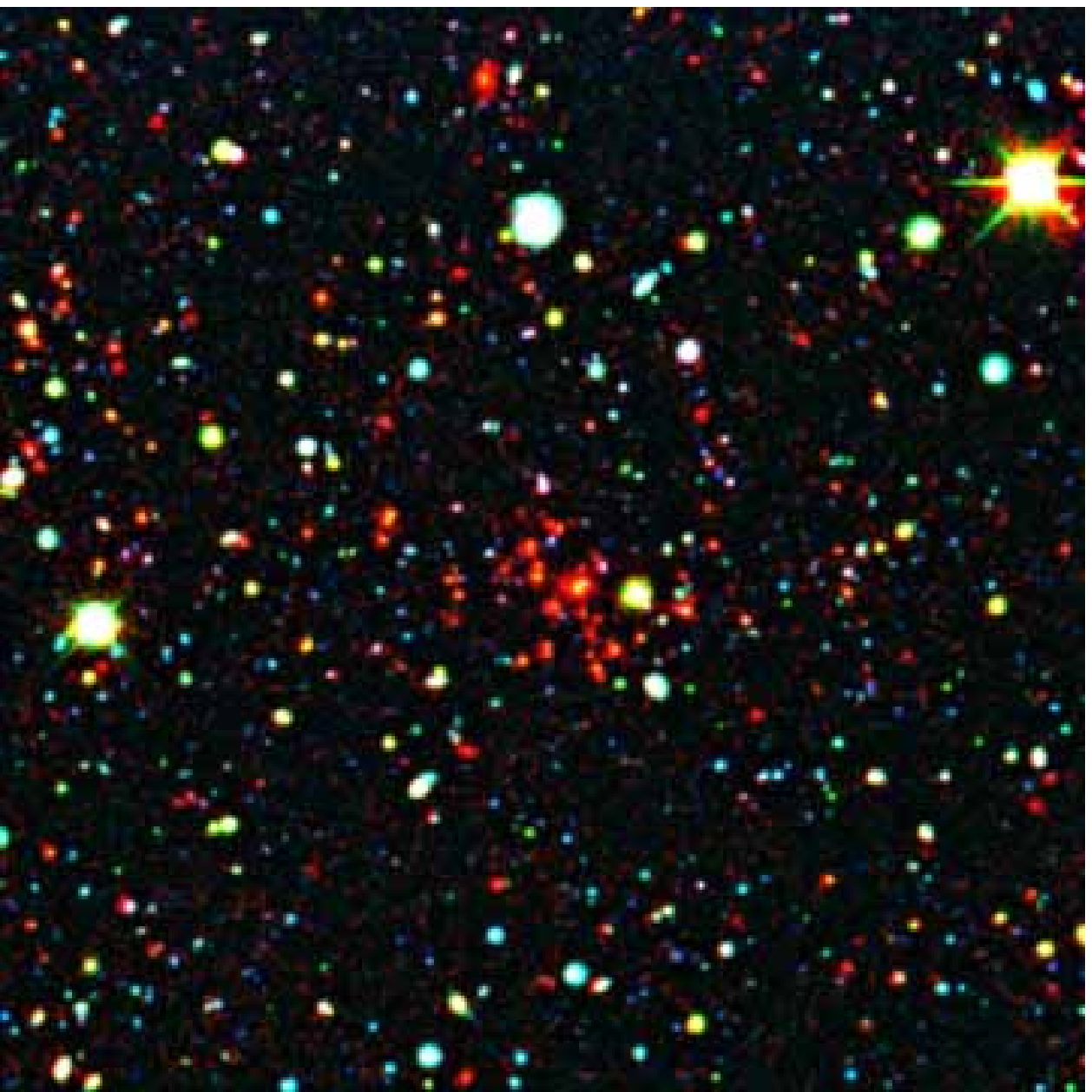}
\caption{
As for Figure \ref{color10.152}, but for 
cluster ISCS J1429.3+3437 at $\left<z_{\rm sp}\right>=1.258$.}
\label{color10.30}
\end{figure}

\begin{figure}[bthp]
\plotone{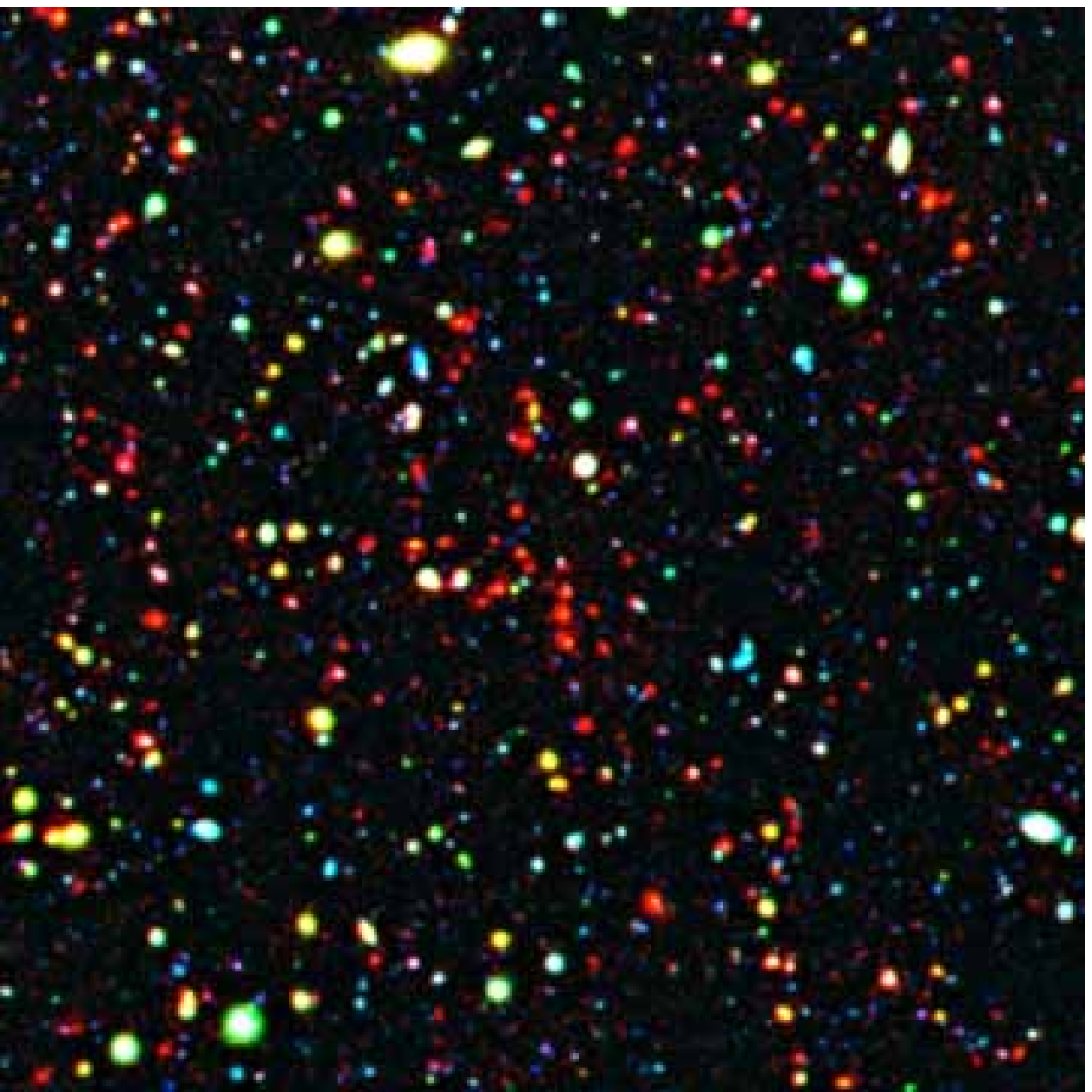}
\caption{
As for Figure \ref{color10.152}, but for 
cluster ISCS J1432.6+3436 at $\left<z_{\rm sp}\right>=1.347$.}
\label{color10.29}
\end{figure}

\begin{figure}[bthp]
\plotone{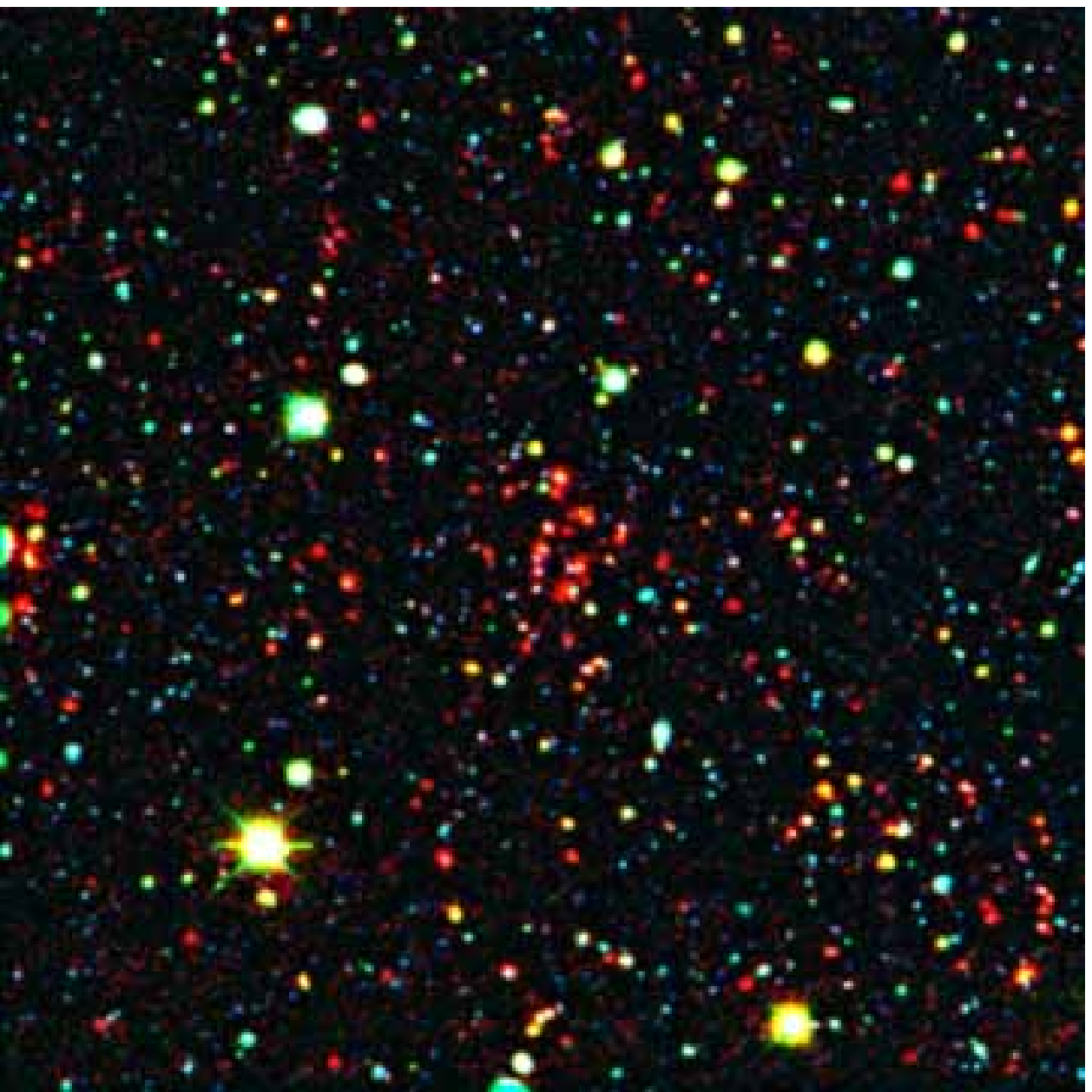}
\caption{
As for Figure \ref{color10.152}, but for 
cluster ISCS J1434.7+3519 at $\left<z_{\rm sp}\right>=1.373$.}
\label{color10.25}
\end{figure}

\begin{figure}[bthp]
\plotone{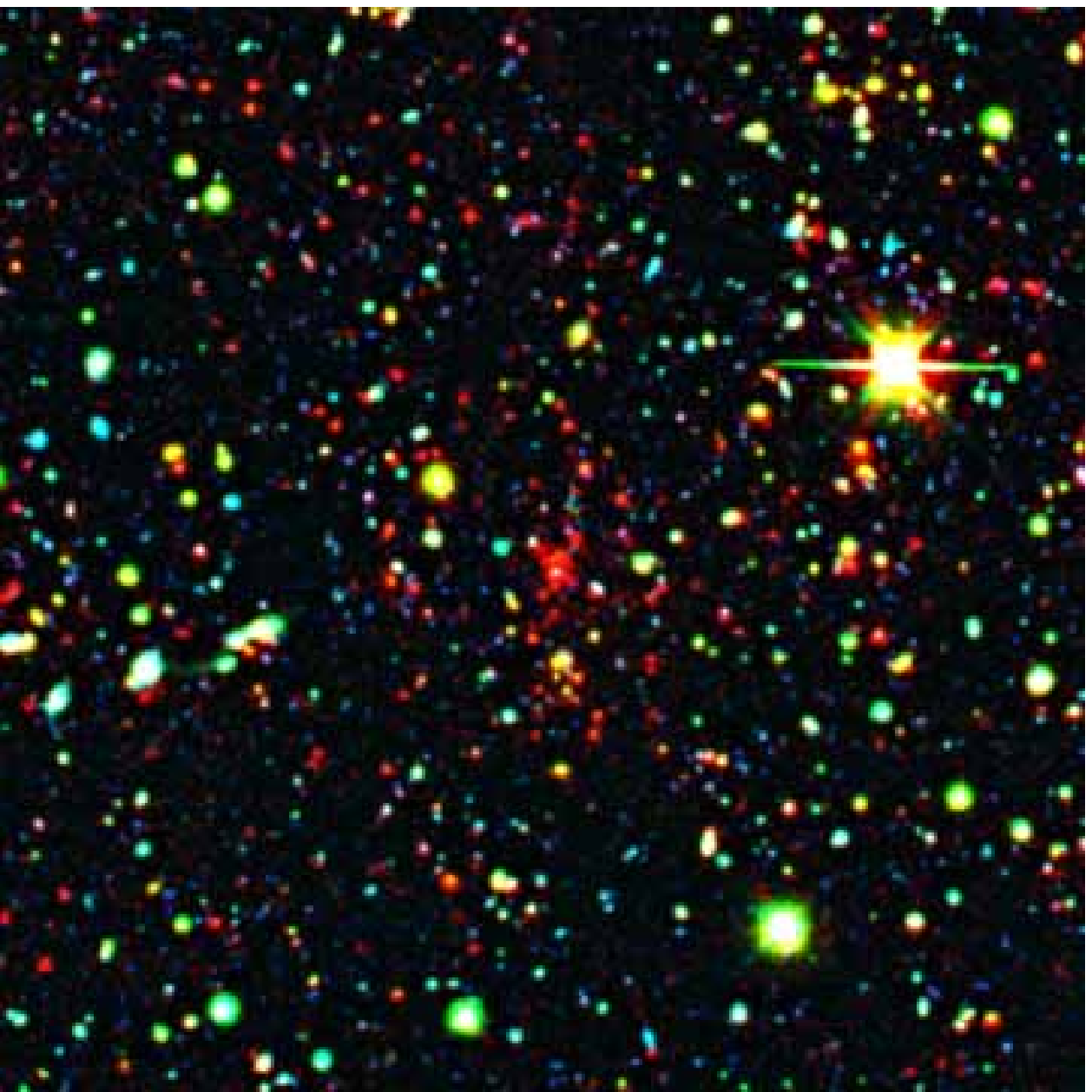}
\caption{
As for Figure \ref{color10.152}, but for 
cluster ISCS J1438.1+3414 at $\left<z_{\rm sp}\right>=1.413$.}
\label{color10.22}
\end{figure}

\subsection{AGES Spectroscopy}
\label{sec:AGES}
 
The extensive AGES spectroscopic database
(C. Kochanek et al. in preparation) can be used 
to spectroscopically confirm clusters at $z < 0.5$,
a redshift at which $L^*$ corresponds to $I = 20$ (Figure 1), 
the magnitude limit for AGES spectroscopy. 
Note that while AGES is 94\% complete to  $I = 18.5$, 
sparse sampling is used for fainter galaxies, and currently
spectra are available for $\sim40\%$ of galaxies
with $18.5 < I < 20$.   
Because of the high surface density of AGES redshifts and
the larger angular scale at $z < 0.5$, and the fact that the 
[4.5] flux limit samples
substantially less luminous galaxies at these redshifts,
two additions to the criteria for
spectroscopic confirmation used at $z > 1$ were imposed.  
To contribute to spectroscopic confirmation for cluster candidates with
$z_{\rm est} < 0.5$, 
galaxies were required to be more luminous
than L* + 1 in [4.5] for a red spike evolving model. 
Also, the average surface density of AGES galaxies over the \boot~field
which would satisfy our high redshift confirmation criteria was calculated
as a function of redshift, and 5 galaxies
above this field level were required for a cluster to be confirmed.  

Of the 335 cluster candidates, 80 have $z_{\rm est} < 0.5$.  
However seven lie near the edges of the \boot~field
and hence were not well observed in AGES.
Of the remaining 73, 61 candidates are confirmed by AGES spectroscopy
using the criteria just described.  This
criterion is perhaps overly stringent: it rejects two clusters 
at $z=0.2$ with 17 or more matching AGES redshifts, but with only 4 
(rather than 5) for galaxies brighter than L* + 1 after field correction. 
All of the other 10 candidates observed by AGES which 
do not meet the confirmation criteria have $z_{\rm est} \ge 0.37$, 
and half have $z_{\rm est} \ge 0.45$.
 
Given the sparse sampling of AGES at $I > 18.5$, which corresponds
to $L^*$ at $z > 0.3$ (Figure 1), 
we believe this confirmation rate validates our estimate that only
$\sim 10\%$ of our cluster candidates arise by chance or due to projection
effects.
Further details on $z < 1$ clusters will be provided in A. Gonzalez et al. 
(in preparation).

\subsection{Keck Spectroscopy}
\label{sec:Keck}

\begin{figure}
\begin{center}
\rotatebox{-90}{\resizebox{0.7\textwidth}{!}{\includegraphics{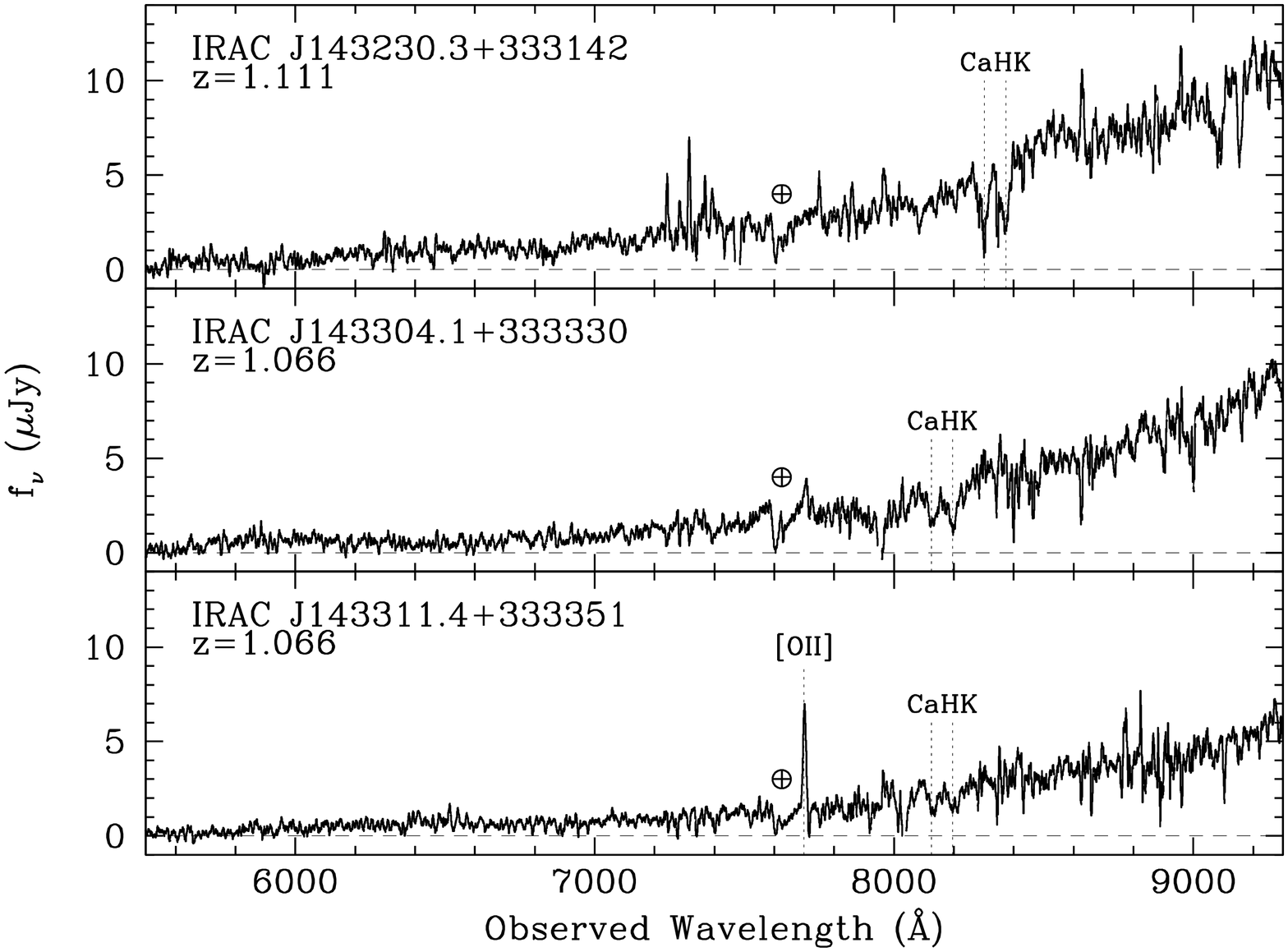}}}
\caption{Optical spectra of three $z>1$ members of IRAC Shallow Survey clusters.  The spectra
were obtained with Keck/DEIMOS in April 2007 and flux calibrated using
an archival sensitivity function.  All three sources have class A spectra (\S \ref{clusterz})
and show clear Ca H and K absorption, typical of early-type galaxies.  
The bottom source also shows strong [\ion{O}{2}]
emission, indicative of star formation.  All three sources have $I \sim 21.7$.
Imperfect sky subtraction is evident in the top spectrum near 7300 \AA, and
the 7600 \AA\ atmospheric A-band is marked in all 3 panels.
\label{spectra}
}
\end{center}
\end{figure}

Most of the high redshift spectroscopic confirmation of ISCS clusters
has been obtained at Keck Observatory.  Three clusters observed
with Keck have been reported previously: ISCS~J1432.4+3332 ($z =
1.112$), ISCS~J1434.5+3427 ($z = 1.243$), and ISCS~J1438.1+3414 ($z =
1.413$) are presented in \citet{Elston2006}, \citet{Brodwin2006}, and
\citet{Stanford2005}, respectively.  Here we provide spectroscopic
confirmation for an additional nine clusters at $z > 1$, as well as
some new spectroscopic information for the initial three clusters.
Table~\ref{SpecObsTable} details new observations of ISCS clusters.
Table~\ref{SpecTable} provides properties of previously unreported,
spectroscopically confirmed cluster members, in declination order for IRAC
sources, followed by serendipitous sources.
Figure \ref{spectra} shows three example spectra obtained in April
2007 with Keck/DEIMOS.

\begin{deluxetable}{lcrcccccccc}
\tabletypesize{\normalsize}
\tablecaption{Keck and Subaru Observations of $z>1$ ISCS Galaxy Clusters
\label{SpecObsTable}}
\tablewidth{0pt}
\tablehead{
\colhead{Cluster} &
\colhead{$\left<z_{\rm sp}\right>$} &
\colhead{Inst.} &
\colhead{UT Date} &
\colhead{Exp. Time (s)} &
\colhead{Conditions}}
\rotate
\startdata
ISCS\_J1434.1+3328 & 1.057 & DEIMOS & 2007 Apr 18$-$19 & 6$\times$1800 & clear; 0\farcs8 - 2\farcs0 \\
ISCS\_J1429.2+3357 & 1.058 & DEIMOS & 2006 Apr 26 & 4$\times$1500 & not photometric \\
ISCS\_J1433.1+3334 & 1.070 & DEIMOS & 2007 Apr 19 & 3$\times$1200 & clear; 0\farcs8 \\
ISCS\_J1433.2+3324 & 1.096 & DEIMOS & 2007 Apr 18$-$19 & 6$\times$1800 & clear; 0\farcs8 - 2\farcs0 \\
ISCS\_J1432.4+3332 & 1.112 &   LRIS & 2005 Jun 03 & 5$\times$1800 & clear; 0\farcs9 (Elston et al. 2006) \\
&                          &  FOCAS & 2006 Apr 21 & 5$\times$1200 & \\
&                          & DEIMOS & 2007 Apr 19 & 3$\times$1200 & clear; 0\farcs8 \\
ISCS\_J1426.1+3403 & 1.135 &   LRIS & 2007 May 19 & 3$\times$1800 & clear; 0\farcs9  \\
ISCS\_J1426.5+3339 & 1.161 &   LRIS & 2006 Apr 04 & 4$\times$1800 & clear \\
ISCS\_J1434.5+3427 & 1.243 &  FOCAS & 2006 Apr 22 & 5$\times$1200 & \\
&                          & DEIMOS & 2007 Apr 18 & 8$\times$1800 & likely cirrus; 1\farcs0 - 1\farcs5 \\
ISCS\_J1429.3+3437 & 1.258 &   LRIS & 2006 Apr 05 & 3$\times$1200 & \\
ISCS\_J1432.6+3436 & 1.347 &   LRIS & 2007 May 21 & 7$\times$1800 & mostly clear; 0\farcs8 \\
ISCS\_J1434.7+3519 & 1.373 &   LRIS & 2005 Jun 02 & 7$\times$1800 & \\
ISCS\_J1438.1+3414 & 1.413 &  FOCAS & 2006 Jun 28 & 9$\times$1200 & \\
&                          & DEIMOS & 2007 Apr 19 & 7$\times$1800 & clear; 0\farcs9 \\
\enddata
\tablecaption{Note that multiple clusters were sometimes observed on a single DEIMOS 
mask.  Thus, the observations listed above are not all independent.}
\end{deluxetable}

\subsubsection{LRIS Observations}

We obtained deep optical slitmask spectroscopy for several clusters using
the dual-beam Low Resolution Imaging Spectrograph \citep[LRIS;][]{Oke1995}
on the 10~m Keck~I telescope during three observing runs between 2005
and 2007.  Slitmasks generally included approximately 15 objects with
photometric redshifts consistent with cluster membership and within
4 arcmin of the nominal cluster centers.  Additional IRAC 4.5$\mu$m
selected sources were included to fill out the slitmasks.  Slitlets had
widths of 1.3 arcsec and minimum lengths of 10 arcsec.  We employed the
D580 dichroic which splits the light at $\sim 5800$ \AA\ between the two
channels of LRIS.  On the red side, the 400 line grating, blazed at 8500
\AA, was used to cover a nominal wavelength range of 5800 to 9800 \AA,
varying somewhat depending on the position of a slit in the mask.  For objects filling
a slitlet, the spectral resolution for this instrument configuration
is $\sim 9$ \AA\ ($R \sim 900$), as determined from arc lamp spectra.
On the blue side, the 400 line grism, blazed at 3400 \AA, provided
coverage from the atmospheric cutoff ($\sim 3200$ \AA) up to the dichroic
cut off.  For objects filling a slitlet, the spectral resolution for this
instrument configuration is $\sim 8$ \AA\ ($R \sim 450$). 

We obtained multiple exposures for each mask, usually with 1800~s per
individual exposure.
Table~\ref{SpecObsTable} details the exposure times and observing
conditions.  The observations were carried out with the slitlets aligned
close to the parallactic angle, and objects were shifted along the long
axis of the slitlets between exposures to enable better sky subtraction
and fringe correction.  

The slitmask data were separated into individual spectra and then reduced
using standard longslit techniques.  The multiple exposures for each
slitlet were reduced separately and then coadded.  The spectra were
reduced both without and with a fringe correction; the former tends to
yield higher quality object spectra at the shorter wavelengths, while
the latter is necessary at the longer wavelengths.  Calibrations were
obtained from arc lamp exposures taken immediately after the object
exposures for the red side, and from arc lamp exposures taken during
the afternoon for the blue side.  Corrections for small offsets in the
wavelength calibration were obtained by inspection of the positions of
sky lines in the object spectra.  Using longslit observations of the
standard stars from \citet{Massey90} obtained during the same observing
runs, we achieved relative flux calibration of the spectroscopy.
While slit losses for resolved sources preclude absolute spectrophotometry 
from the slit mask
data, the relative calibration of the spectral shapes should be accurate.
One--dimensional spectra were extracted from the sum of all the reduced
data for each slitlet for both the red and blue sides.  For the targets
in the high-redshift clusters, generally only the red side data proved
useful.

\subsubsection{DEIMOS Observations}

Additional spectroscopy was obtained with the Deep Imaging Multi-Object
Spectrograph \citep[DEIMOS;][]{Faber2003} on the 10~m Keck~II telescope,
a second generation instrument with significantly more multiplexing
capabilities as compared to LRIS, albeit without the blue sensitivity.
During an observing run in April 2006, we targeted Bo\"otes active
galaxy candidates selected on the basis of mid-infrared colors
\citep[e.g.,][]{Stern2005}, but included candidate cluster members in
one mask.  In April 2007, while observing host galaxies of high redshift
cluster supernovae (Perlmutter et al., in preparation), we also observed
candidate cluster members, typically with more than one high redshift
cluster candidate observed in each wide area mask.

For both observing runs, the 600ZD grating ($\lambda_{\rm blaze}
= 7500$ \AA; $\Delta \lambda_{\rm FWHM} = 3.7$ \AA) and a GG455
order-blocking filter were used.  DEIMOS data were processed using a
slightly modified version of the pipeline developed by the DEEP2 team at
UC-Berkeley\footnote{\tt \url{http://astro.berkeley.edu/$\sim$cooper/deep/spec2d/}}.
Although neither run was completely photometric, relative
flux calibration was achieved from observations of standard stars from
\citet{Massey90}.

\subsection{Subaru Spectroscopy}\label{sec:Subaru}

Followup spectroscopic observations were also obtained 
with the FOCAS spectrograph \citep{Kashikawa2002} 
on the 8.2-m Subaru telescope in April and June 2006.
For these observations, we used the instrument in multi-object
spectroscopy mode with 0.8 arcsec width slits, 300R grism, and
SO58 order-sorting filter, which provided about 15 spectra
from 5800\AA\ to 10000\AA\ with a spectral resolution $R\sim500$.
Total exposure times were 2-6 hours. The FOCAS data were reduced
with IRAF using standard methods including a fringe correction.
Wavelength calibration was done using OH airglow emission lines.
Absolute flux was calibrated using standard star (Feige 34,
Wolf 1346, Hz 44) spectra taken during the same nights.

\subsection{Notes on Individual Clusters}
\label{sec:Notes}

\subsubsection{ISCS~J1434.1+3328 (z = 1.057)\label{1434.1+3328}}

ISCS~J1434.1+3328 was observed with DEIMOS on a mask optimized for
observing a high redshift supernova identified in a different ISCS
high redshift cluster (Perlmutter et al., in preparation) which has not yet
been spectroscopically confirmed.  Four candidate
members of ISCS~J1434.1+3328 were targeted, and two more spectroscopic
members were identified from additional slitlets targeting
4.5~$\mu$m selected sources to fill the mask.  All six confirmed
members have clearly identified Ca H and K absorption lines and
D4000 breaks, and none of the galaxies exhibit emission features in the
wavelengths covered by the DEIMOS observations ($\approx 5500 -
9500$~\AA).

\subsubsection{ISCS~J1429.2+3357 (z = 1.058)\label{1429.2+3357}}

ISCS~J1429.2+3357 was observed with DEIMOS, on a slitmask
optimized for observing faint, mid-infrared selected AGN candidates
\citep[e.g.,][]{Stern2005}.  Eight candidate cluster members were included
in the slitmask, six of which were confirmed spectroscopically to reside
at $z \approx 1.06$.  Only the brightest galaxy, IRAC~J142912.9+335808,
shows [\ion{O}{2}] emission; the remaining redshifts were derived on the
basis of Ca H and K absorption lines and/or D4000 breaks.  In addition,
one galaxy targeted spectroscopically as an IRAC-selected AGN candidate,
IRAC~J142916.1+335537, is a cluster member, bringing the tally
to seven spectroscopically confirmed cluster members.  The spectrum of
this source shows strong, narrow (400 km\ s$^{-1}$) emission lines from
[\ion{Ne}{5}] $\lambda 3426$, indicating that it is indeed an active
galaxy.

\subsubsection{ISCS~J1433.1+3334 (z = 1.070)\label{1433.1+3334}}

ISCS~J1433.1+3334 was observed with DEIMOS on a mask optimized for
observing a high redshift supernova identified in ISCS~J1432.4+3332 
(Perlmutter et al., in preparation).  Eight candidate
members of ISCS~J1434.1+3328 were targeted, of which six were
confirmed as cluster members, one was found to be slightly foreground
to the cluster, and one was slightly behind the cluster.  All eight
candidates show clear D4000 breaks.  Many additional cluster members
were identified on this mask, either serendipitously or as
targeted IRAC-selected, $z > 1$ galaxies, bringing the total 
number of spectroscopically confirmed cluster members to 20.   
As seen in the bottom two panels of Figure \ref{spectra}, 
the confirmed sources show a range of spectral properties.
While most show Ca H and K absorption and D4000 breaks,  some also show
emission features likely due to either star formation or AGN activity.

\subsubsection{ISCS~J1433.2+3324 (z = 1.096)\label{1433.2+3324}}

ISCS~J1433.2+3324 was observed with DEIMOS on the same mask as
ISCS~J1434.1+3328.  Five candidate
members of ISCS~J1434.1+3328 were targeted, of which two were
confirmed as cluster members, one was found to be foreground, and
two yielded inconclusive spectra.  Four additional cluster members
were identified in the same mask, from IRAC-selected sources.
The two confirmed cluster members which were specifically targeted
show strong Ca H and K absorption and lack emission lines.  The
other four confirmed members all show [\ion{O}{2}] emission, with
three of the four also showing D4000 breaks.

\subsubsection{ISCS~J1432.4+3332 (z = 1.112)\label{1432.4+3332}}

The spectroscopic observations which confirmed ISCS~J1432.4+3332
at $z = 1.11$ are described in \citet{Elston2006}, but no data on
individual sources was presented there.  The \citet{Elston2006}
result was based on nine spectroscopically confirmed cluster members, two of
which were selected not as cluster members, but rather as mid-IR
selected AGN,  and that data is now included in Table~\ref{SpecTable} of this
paper.  One of the candidate members of this cluster hosted a
high-redshift supernova (Perlmutter et al., in preparation) and was thus
targeted for additional DEIMOS and FOCAS slitmask spectroscopy.  
An example DEIMOS spectrum for this cluster is shown in the top panel 
of Figure \ref{spectra}.  In total, there are now 23 spectroscopically 
confirmed members in the cluster.

\subsubsection{ISCS~J1426.1+3403 (z = 1.135)\label{1426.1+3403}}

Seven candidate members of this cluster were confirmed spectroscopically
during our LRIS observations in May 2007.  All seven galaxies show red
continuum emission and/or a clear D4000 break.  Two of the sources
also show [\ion{O}{2}] emission.

\subsubsection{ISCS~J1426.5+3339 (z = 1.161)\label{1426.5+3339}}

Four of the five spectroscopically confirmed members of this cluster show
[\ion{O}{2}] emission, a higher proportion than is typical for this program.
This is possibly a selection effect; sources with line emission are the
easiest to spectroscopically confirm.  All five confirmed members show
breaks typical of early-type galaxies (e.g., D2900 and/or D4000).

\subsubsection{ISCS~J1434.5+3427 (z = 1.243)\label{1434.5+3427}}

This cluster is discussed in \citet{Brodwin2006}, where eight
spectroscopic members were presented.  Table~\ref{SpecTable} 
of the current paper does not duplicate those data, and instead lists
three additional spectroscopic members that have since been identified.

\subsubsection{ISCS~J1429.3+3437 (z = 1.258)\label{1429.3+3437}}

This cluster has nine spectroscopically confirmed members.  One is an
optically-bright AGN from the AGES survey, while the rest were confirmed
spectroscopically by LRIS. Four of the Keck/LRIS sample
show [\ion{O}{2}] emission, while the other four show only spectral
breaks and absorption lines.

\subsubsection{ISCS~J1432.6+3436 (z = 1.347)\label{1432.6+3436}}

This cluster has eight spectroscopically confirmed members, all
from our Keck/LRIS observations in May 2007.  Only one
of the sources shows [\ion{O}{2}] emission; the rest of the redshifts
are on the basis of continuum breaks at 2640 and 2900 \AA, as well
as \ion{Mg}{2} $\lambda$ 2800 absorption.

\subsubsection{ISCS~J1434.7+3519 (z = 1.373)\label{1434.7+3519}}

This cluster has five confirmed members, one of which was serendipitously
identified in the spectroscopy, all from our Keck/LRIS spectroscopy in
June 2005.
Three of the members show [\ion{O}{2}], and the other two show spectral
breaks characteristic of early-type galaxy spectra.

\subsubsection{ISCS~J1438.1+3414 (z = 1.413)\label{1438.1+3414}}

This cluster was first published in \citet{Stanford2005}, 
at which time it was the highest redshift galaxy cluster known.  Two
candidate cluster members hosted supernovae in the {\it HST}/ACS
program of Perlmutter et al. (in preparation), and this field thus has
been the target of additional spectroscopy from Subaru and Keck.
Five new cluster members have been confirmed, listed in Table~\ref{SpecTable}.
Data on the original five members is not duplicated from \citet{Stanford2005}.
Of the five new members, one is an [\ion{O}{2}] emitter, serendipitously
identified in a slitlet targeting another source.  Three show only
Ca H and K absorption lines, with no emission lines identified.
IRAC~J143816.8+341440 (22.3 in Table 3) is an AGN, showing [\ion{Ne}{4}], [\ion{O}{2}],
and [\ion{Ne}{3}] emission lines.

\subsection{Cluster Masses\label{sec:masses}}

With 20 or more spectroscopic redshifts, it is possible to estimate
cluster masses via scaling relations using the velocity dispersion.  
The line of sight velocity
dispersion for the 20 spectroscopic member galaxies in cluster 19
(ISCS~J1433.1+3334) at $z=1.070$ is 760 km s$^{-1}$ in the rest frame, 
and for the 23 spectroscopic member galaxies in cluster 17
(ISCS~J1432.4+3332) at $z=1.112$ it is 734 km s$^{-1}$.

For these two clusters, which are among the richest in the sample, 
the velocity dispersion of $\sim750$ km s$^{-1}$ corresponds 
to a virial mass of $\sim 3.8~{\rm r_v} \times 10^{14} M_\odot$
where ${\rm r_v}$ is the virial radius in Mpc
\citep[e.g. equation 4 of][]{Carlberg1996}.
The x-ray temperature corresponding to $\sigma=750$ km s$^{-1}$ is 
3.9 keV \citep[Table 4 of][]{XueWu2000}, which in turn
gives a mass of $\sim 4 \times 10^{14} M_\odot$ \citep{Shimizu2003}.  
This is consistent with the average halo mass of 
$\sim 10^{14}M_\odot$ estimated by \citet{Brodwin2007} for the sample.

Stellar luminosities can also be used to make a rough estimate of cluster masses,
scaling to the Coma cluster via the red spike model.
Within the $650\times850$ kpc region of Coma sampled by 
\citet{Eisenhardt2007} the integrated $K$-band luminosity to $L^*+1$
is $54 L^*$, slightly higher than the values shown in column 11 of Table \ref{z1table},
but in a significantly smaller effective radius of 0.42 Mpc.
In their Figure 1, Geller, Diaferio, \& Kurtz (1999) 
show a mass for the Coma cluster of 
$\sim 3.5 \times 10^{14} M_\odot$ within this radius (for h=0.7).  At r=1 Mpc 
they show about double that mass, and suggest the total mass for Coma is about double this 
again.  Hence if the profile and $M/L_K$ for our clusters is similar to Coma, 
allowing for red spike model evolution in $L_K$, 
the $L_{\rm tot}/L^*$ values in 1 Mpc radius provided in Table 1 
scale to total cluster masses of
$\sim 1 - 6 \times 10^{14} M_\odot$ (i.e. $\sim 0.1 - 0.4 M_{\rm Coma}$).
Clusters 17 and 19 are at the top of this range, again providing mass estimates
consistent with those found using the velocity dispersion. 
More detailed exploration of this approach will
require photometry which takes account of the extent to which  
a 5\arcsec\ diameter aperture fails to include 
all of the light from such galaxies, or includes
light from multiple galaxies 
(as determined from higher spatial resolution imaging).

\section{Discussion\label{sec:discussion}}

As noted in the introduction, out to $z \sim 1$ 
substantial evidence exists which is consistent with an extremely
simple formation history for cluster galaxies, in which
their stars are formed in a short burst at high redshift, and
they evolve quiescently thereafter (ie a ``red spike'' model).   
The colors of luminous cluster galaxies
out to $z \sim 1$ typically fall on a tight sequence which is red relative
to field galaxies at similar redshift (the ``red sequence''), with a mean color
which evolves as the red spike model predicts.   

The most luminous red sequence galaxies are the reddest, a
correlation which is attributed to higher metallicity in higher
mass galaxies (\citealt{KodamaArimoto97}; but see also \citealt{FerrerasSilk2003}).  
This correlation (i.e the color-magnitude or mass-metallicity relation) is 
explained by additional cycles of star formation and enrichment
in more massive galaxies, as they are able to retain their gas
more effectively against supernovae-driven winds \citep{ArimotoYoshii87}.
In the red spike formation paradigm, this is a natural consequence.  
In the context of the hierarchical merging galaxy formation models, 
a unique correlation between stellar mass and metallicity is a greater challenge.
While the inclusion of feedback (whether by supernovae or by AGN) 
in hierarchical models stops the buildup of galaxy mass  
\citep[e.g.][]{KauffmannCharlot98,  NagashimaYoshii2004},
such models have difficulty reproducing 
the exact color and slope of the color-magnitude relation in
clusters over the full redshift range for which it has been measured.
Recent work shows promise, however, as feedback behavior and other
``gastrophysical'' effects are taken more into account
\citep[e.g.][]{deLucia2006}.  

If the onset of the star formation ``spike" is simultaneous
for all cluster galaxies, and if the color-magnitude relation is caused by 
more protracted spikes in more massive galaxies, 
then as one approaches the  star-forming epoch, massive
galaxies might be expected to eventually become bluer, reversing the slope
of the color-magnitude relation.  In fact no measurable change is seen
in color-magnitude slope out to $z \sim 1$ \citep[e.g][]{SED98, Mei2006},
which suggests either formation redshifts well before $z = 1$, or 
that the spikes begin earlier in more massive galaxies, perhaps
ending rather than beginning simultaneously.  
The small and unchanging scatter of the red sequence in clusters out to $z\sim1$
\citep[e.g.][]{SED98, Blakeslee2003, Tran2007} 
also argues for synchronized or very early spikes and 
against a primarily age-based origin of the
color-magnitude relation, since the scatter would increase by a much
larger amount, and faster, than has been observed.

Testing to what extent red spike models remain consistent with cluster galaxy
data at $z > 1$ is one of the primary motivations for the present study.
We begin by constructing the color-magnitude relations for the 12
spectroscopically confirmed clusters in Table \ref{z1table}.

\subsection{Color-Magnitude Diagrams\label{sec:cmds}}

\begin{figure}
\plotone{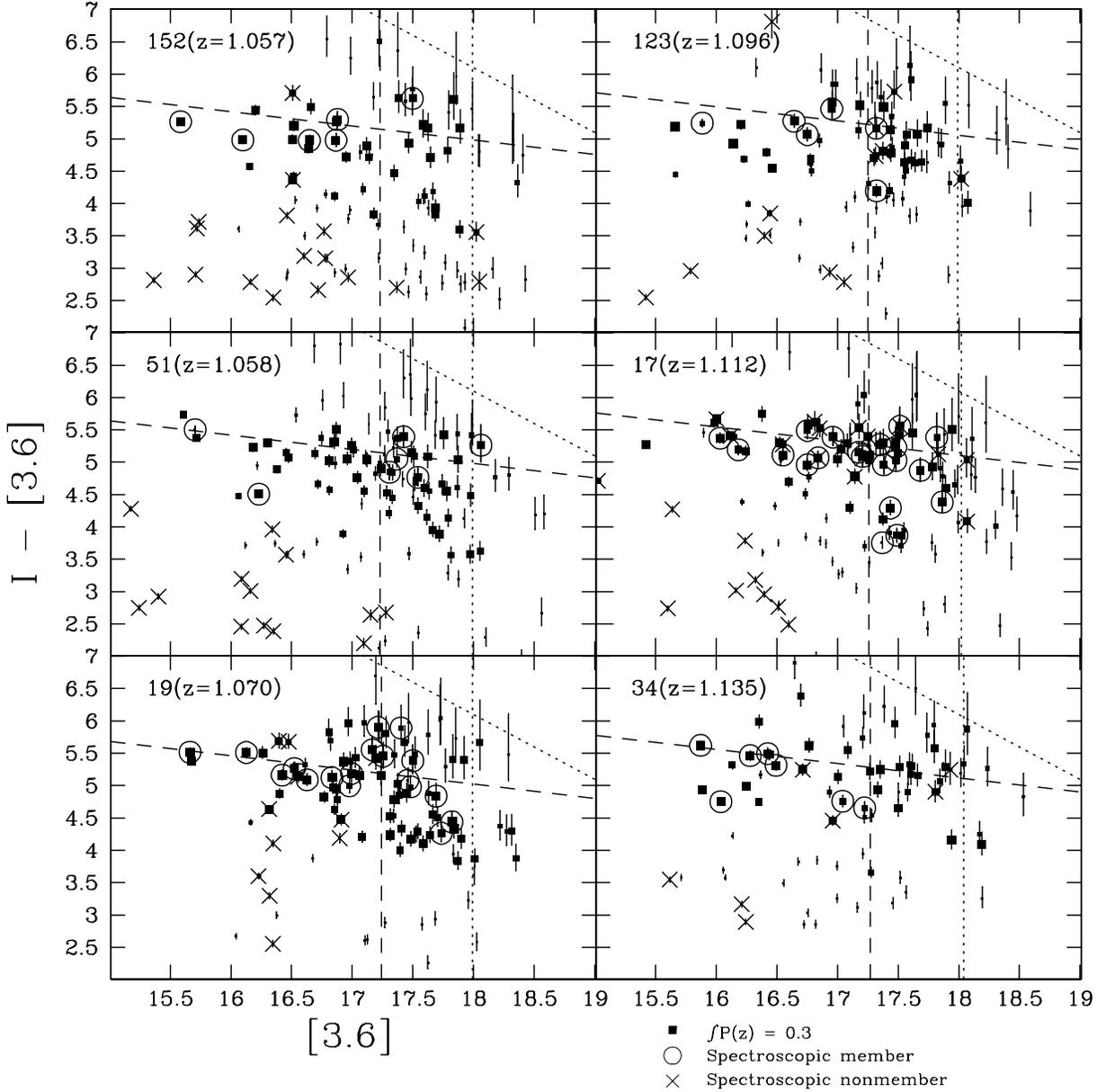}
\caption{
Observed $I - [3.6]$ vs. [3.6] for galaxies within 1 Mpc of the center of
the spectroscopically confirmed clusters at $z > 1$, in order of 
increasing redshift.  The catalog number and spectroscopic redshift for each
cluster from Table \ref{z1table} 
are given in the upper left corner of each sub-panel.
The symbol area is proportional to the
galaxy's integrated redshift probability in the range
$z_{\rm est}\pm{0.06(1+z_{\rm est})}$.  A symbol with an integrated probability of 0.3 is
shown for reference in the lower right portion of the
figure.  Additional details are provided in \S \ref{sec:cmds}.
\label{CMDs1}
}
\end{figure}

\begin{figure}
\plotone{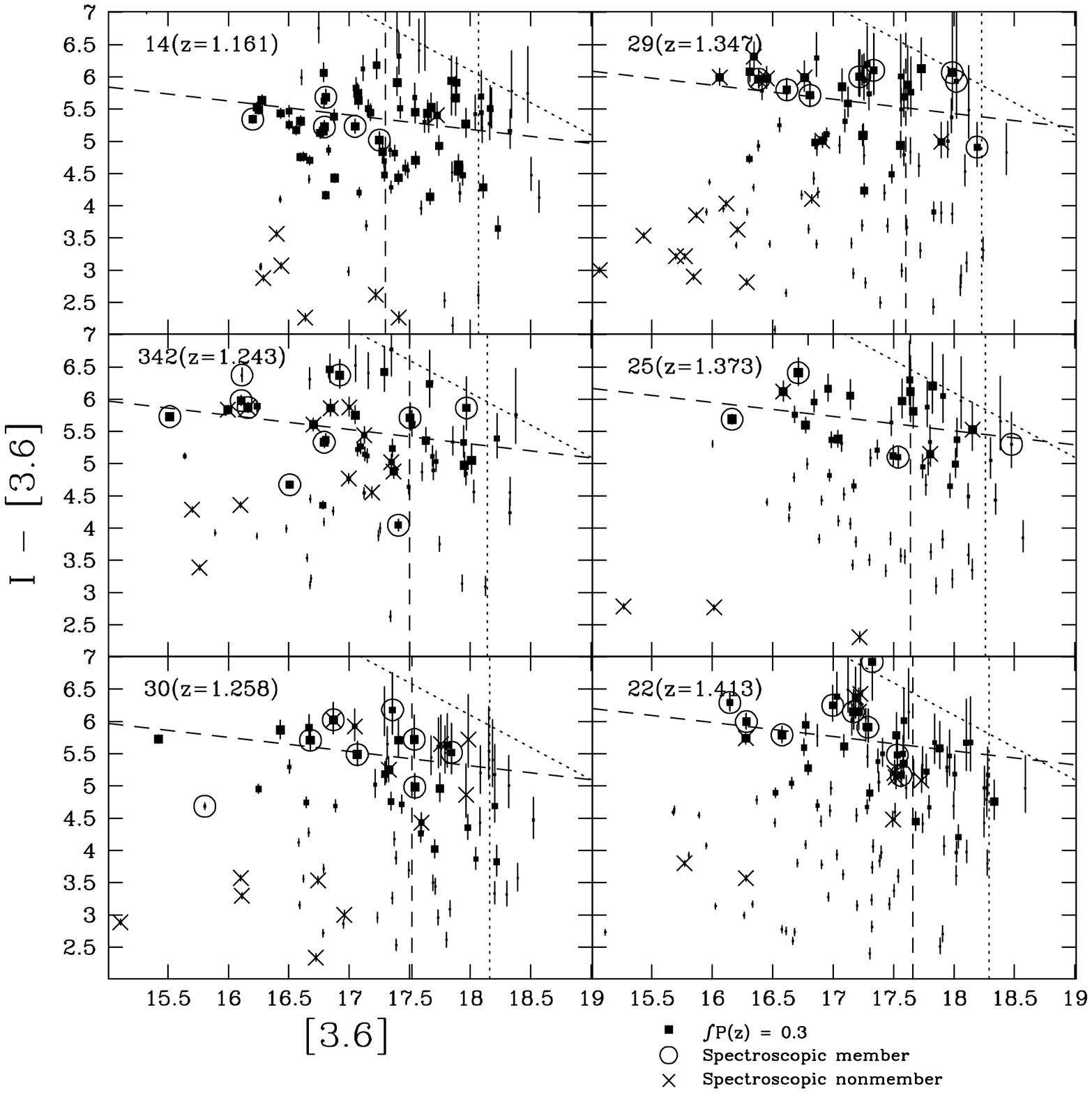}
\caption{Observed $I - [3.6]$ vs. [3.6] for the remainder of
the spectroscopically confirmed clusters at $z > 1$, in order of 
increasing redshift. See figure \ref{CMDs1} for details.
\label{CMDs2}
}
\end{figure}

Figures \ref{CMDs1} and \ref{CMDs2}
present color magnitude diagrams for the clusters listed in Table \ref{z1table}.  The $I - [3.6]$ color was selected because 
these filters bracket the
4000\AA\ break most tightly at $z > 1$.  
The symbol area is proportional to the integral of the 
object's redshift probability distribution over the range 
$z_{\rm est} \pm 0.06(1+z_{\rm est})$,
which is the rms dispersion in individual photometric
redshifts \citep{Brodwin2006}.  Circled symbols are spectroscopically
confirmed members, while crosses indicate objects known 
{\em not} to be members on the basis of spectroscopy.   
Figures \ref{CMDs1} and \ref{CMDs2} include
objects within 1 Mpc of the cluster center,
with $> 5\sigma$ detections in both [3.6] and [4.5],   
and with $> 2\sigma$ detections in $I$; or with spectroscopic redshifts. 
As noted in \S \ref{sec:phot-limits}, the $5\sigma$ limit in [3.6] is 18.6 mag,
so the limiting factor in object selection for Figures \ref{CMDs1} 
and \ref{CMDs2} is from
the $5\sigma$ limit of 17.8 mag in [4.5].  
We have used the  0.1 Gyr burst, $z_f = 3$ red spike model 
shown in Figure \ref{m_vs_z} 
to calculate the corresponding [3.6] mag as a function of redshift, 
and this is shown in Figures \ref{CMDs1} and \ref{CMDs2} 
by the vertical dotted lines. 
The $I$ limit is shown by the diagonal dotted lines in Figures 
\ref{CMDs1} and \ref{CMDs2}. 
The vertical dashed line plots the expected L* magnitude at [3.6] 
for the red spike model.
The sloped dashed line shows the observed  
$U - H$ color-magnitude slope of 0.22 for the Coma cluster 
from the data of \citet{Eisenhardt2007}, normalized at the 
red spike model [3.6] magnitude and $I - [3.6]$ color for an L* galaxy.  
In this redshift range observed $I - [3.6]$ is close to rest $U - H$.
Note that while the normalization evolves according to
the red spike model, a constant, {\em unevolving} slope value is used.   

The brightest galaxy likely to be a member
is typically 1 to 2 magnitudes brighter than the expected L* magnitude
for the red spike model (see Table \ref{z1table}).  
It is noteworthy that the brightest galaxy
in the Coma cluster is $\approx 6$ times brighter than L* in the $H$
band \citep{DePropris1998, Eisenhardt2007}, 
suggesting less than a factor of two growth 
in stellar mass in such galaxies since $z \sim 1.5$.  
{\it HST} imaging should be used to assess this more carefully.   

Because these clusters were {\em not} selected 
on the basis of containing a red sequence (\S \ref{sec:detn}),  
the fact that the highest probability cluster members 
tend to track the passively evolving Coma cluster sequence shown by
the dashed line is significant. 
Evidently the red sequence persists 
in dense environments to $z = 1.4$, 
even when not used as a selection criterion.  In addition there is
no indication that the slope of the color-magnitude relation has changed
sign, and in fact a non-evolving Coma cluster slope appears 
consistent with the data.  

Comparing the data for e.g.~cluster 14 (ISCS~J1426.5+3339 at $z=1.16$) 
vs.~29 (ISCS J1432.6+3436 at $z=1.34$) suggests
real differences in the scatter of the color-magnitude relation do exist. 
The presence of luminous galaxies substantially bluer than the red sequence in
several clusters, a number of them spectroscopically confirmed, is also noteworthy.  
These may represent a population of massive star-forming galaxies which fade 
onto the red sequence in rich clusters, but is not found in field surveys
\citep[e.g.][]{Bell2004}.
It should be noted that the very low scatters reported
for $ z > 1$ clusters in e.g. \citet{Blakeslee2003} and 
\citet{Mei2006} are calculated for galaxies known to have
early-type morphologies, using much deeper photometry than
the survey/discovery data used here.
Further investigation of the color-magnitude relations
is deferred to a future paper which will use deeper
{\it Spitzer} and {\it HST} imaging presently
being obtained.

Although the survey data used to identify the clusters does
not enable accurate photometry of individual galaxies in clusters,
we can calculate mean properties for galaxies in each cluster with
some confidence. In the next section we briefly
explore the color evolution and color-magnitude relation 
of the entire cluster sample. 

\subsection{Color vs. Redshift\label{sec:Im3p6_vs_z}}
\label{sec:Im3p6vsz}

\begin{figure}
\plotone{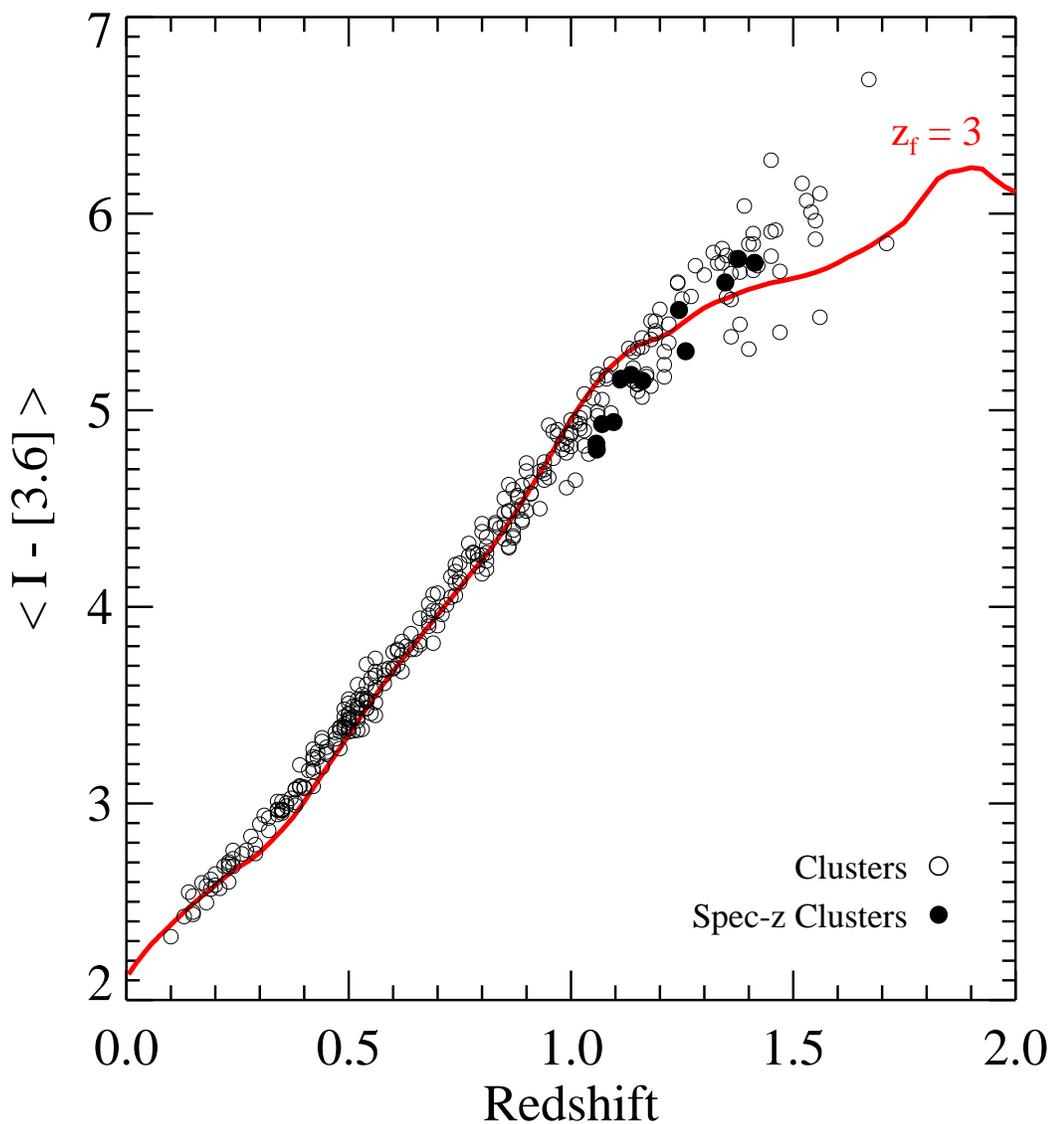}
\caption{
Mean observed $I - [3.6]$ vs estimated cluster redshift $z_{\rm est}$
for galaxies within 1 Mpc of the cluster center 
and with integrated redshift probabilities $> 0.3$ in the range $z_{\rm est}\pm{0.06(1+z_{\rm est})}$.  
The solid red line is the predicted $I - [3.6]$ color for the
red spike model. Filled symbols show the mean $I - [3.6]$ colors for the
spectroscopically confirmed $z > 1$ clusters listed in Table \ref{z1table}.
\label{Im3p6vsz}
}
\end{figure}

Figure \ref{Im3p6vsz} plots the average $I - [3.6]$ color 
for galaxies within 1 Mpc of the 
cluster centers, 
and whose integrated redshift probability distribution in the
range $z_{\rm est} \pm 0.06(1+z_{\rm est})$ exceeds 0.3.  
The mean values are calculated after iteratively clipping $3\sigma$ outliers. 
As implied by the color-magnitude diagrams for the 
spectroscopically confirmed clusters (\S \ref{sec:cmds}), 
the red spike model (0.1 Gyr burst at $z_f = 3$) 
provides a remarkably good fit
to the empirical data, validating the survey design assumptions 
illustrated in Figure \ref{m_vs_z}.  

\begin{figure}
\plotone{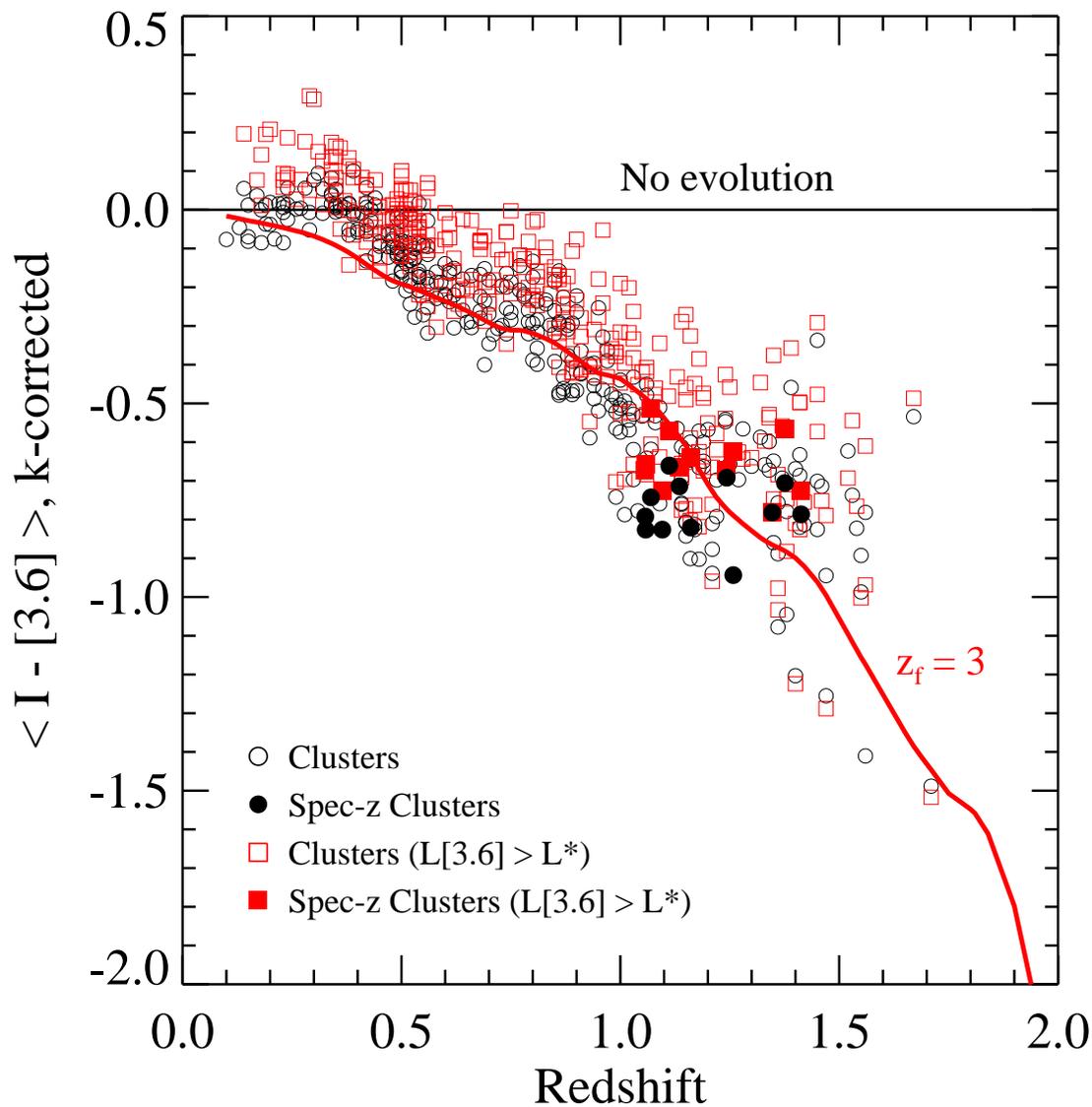}
\caption{
As for Figure \ref{Im3p6vsz}, but with the k-correction subtracted.
Red squares are the mean observed $I - [3.6]$ for galaxies 
brighter than the expected [3.6] magnitude for an L* galaxy for the red spike model.  
The solid red line is the predicted $I - [3.6]$ color for the
red spike model. Filled symbols show the 
mean $I - [3.6]$ colors for the spectroscopically confirmed $z > 1$ clusters 
listed in Table \ref{z1table}. 
\label{Im3p6relNEvsz}
}
\end{figure}

Much of the increase in $I - [3.6]$ with 
redshift is due to the change in the rest frame wavelengths observed, 
and in Figure \ref{Im3p6relNEvsz} we plot the average color
relative to a no-evolution model (i.e. the k-corrected color).  
The no-evolution model is simply the
red spike model at z=0, i.e. at an age of 11.3 Gyr for the stellar
population.  Clearly the evolving red spike model 
is a better fit to the data than is the k-correction alone. 
Note that the photometric redshifts (\S \ref{sec:photz}) are calculated using
{\em non}-evolving templates, so this result is not a foregone
conclusion, particularly at $z > 1$.

The mean $I - [3.6]$ colors in Figure \ref{Im3p6vsz} include galaxies
down to the $13.3\mu$Jy survey limit at $4.5\mu$m, and therefore
less luminous galaxies contribute to the mean at lower redshifts.
The color-magnitude relation shows that less luminous galaxies 
are bluer, and massive galaxies are quite quiescent today, so
a constant flux limit will lead to a systematic bias towards
more luminous, massive galaxies and redder mean
color with increasing redshift.  
On the other hand the excellent fit of the
red spike model implies that a 
constant {\em luminosity} limit for the sample
would select galaxies with smaller stellar masses at high redshift,
since a constant stellar mass, passively evolving, will be more luminous
as one approaches the formation redshift.
To avoid this bias, in Figure \ref{Im3p6relNEvsz}
we also show (with red squares) the mean colors for galaxies 
brighter than the red spike passively-evolving L* in [3.6].

It is evident that the mean color of the more luminous galaxies
is systematically redder, and hence that the color-magnitude relation
continues to hold out to $z \sim 1.5$.  The color offset is 0.1 
magnitudes, independent of redshift.  The persistence of the
essentially unchanged color-magnitude relation slope, with a passively
evolving intercept, out to lookback times 
within 4 Gyr of the Big Bang is a phenomenon that models of cluster
galaxy formation must account for.  The implication is that the
correlation between high stellar population metallicity and
stellar mass is already in place at $z \sim 1.5$, and that the
star formation era remains well in the past for these galaxies.  
 
\begin{figure}
\plotone{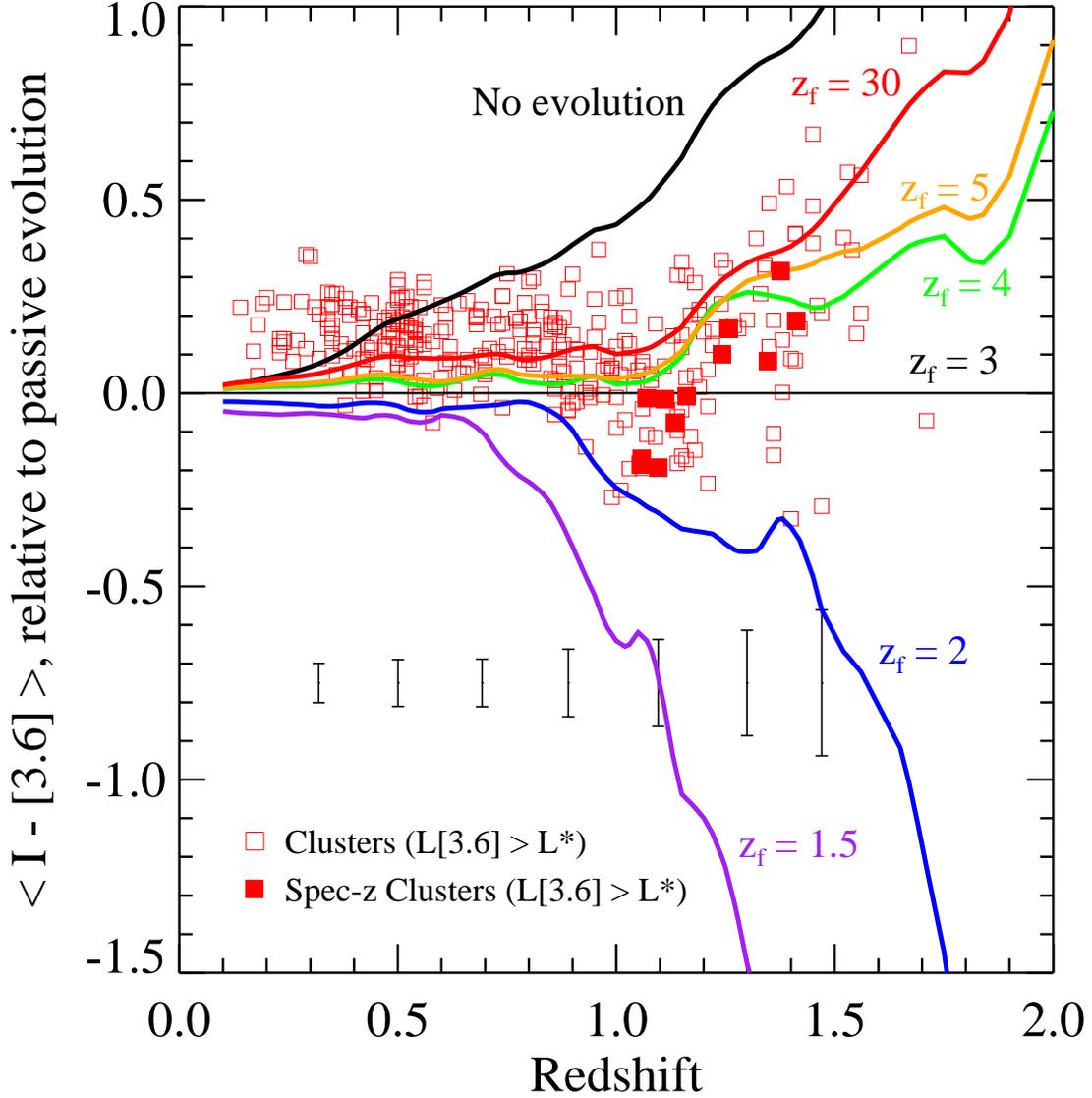}
\caption{
As for Figure \ref{Im3p6vsz}, but with predicted $I - [3.6]$
color for the red spike model subtracted.  Means are shown only for
galaxies brighter than the expected [3.6] magnitude for an L* galaxy 
for the red spike model.  The black horizontal line  is the
0.1 Gyr burst, $z_f=3$ (red spike) model prediction.  The purple,
blue, green, orange, red, and black lines show the expected 
relative $I - [3.6]$ color
for a 0.1 Gyr burst starting at $z_f = 1.5$, 2, 4, 5, 30, and for no evolution (k-correction) respectively.  
Representative values of the uncertainty in each cluster's
mean $I - [3.6]$ color are shown at several redshifts.
Filled symbols show the 
mean $I - [3.6]$ colors for the spectroscopically confirmed $z > 1$ clusters 
listed in Table \ref{z1table}. 
\label{Im3p6relPEvsz}
}
\end{figure}

In Figure \ref{Im3p6relPEvsz} the mean $I - [3.6]$ colors
are plotted with the predicted $I - [3.6]$
color for the red spike model subtracted.
The mean colors are for galaxies more luminous in [3.6] than the red spike L*. 
Note the red spike model
is a good match to $L^*$ galaxy colors in figures \ref{Im3p6vsz} and \ref{Im3p6relNEvsz}, 
so the offset in color out to $z \sim 1$ is attributable to the color-magnitude
relation.  

Other effects may be in play. In Table 1 the $z_{\rm est}$ values tend to be
lower than the $\left<z_{\rm sp}\right>$ values, which may lead to colors appearing
$\sim 0.1$ mag redder relative to the red spike model than if $\left<z_{\rm sp}\right>$
were available for all $z > 1$ clusters.  The average colors
of cluster members may still be redder than shown, however, as no attempt
has been made to correct the average cluster colors for field contamination,
and field galaxies tend to be bluer than cluster galaxies.  Blending issues
could lead to colors which are systematically biased.  Given these uncertainties,
the excellent agreement with the red spike model is the more remarkable.  
Nevertheless, at $z > 1$ the trend is towards
colors which are increasingly redder than the
red spike model, for both the full sample and the spectroscopically confirmed subset, 
though there is clearly a range.   Hence even higher
formation redshifts (as high as $z_f = 30$) are favored for most
$z > 1$ clusters, within the context of red spike models.

\subsection{Clusters at $1.5 < z < 2$\label{sec:z2clusters}}

Figure \ref{Im3p6vsz} shows that
that there are relatively few cluster candidates in the IRAC
sample with $z_{\rm est} > 1.5$. 
Why is this? 
The red spike model fits, 
circumstantial evidence from BCG luminosity, and the persistence of
the color-magnitude relation suggest that the stellar populations were
formed -- and assembled -- well before $z \sim 1.5$.
From Figure \ref{m_vs_z}, such clusters should be detectable
by the IRAC Shallow Survey. 

A potential selection effect against $z > 1.5$ galaxies is that
photometric redshifts in this range have increasingly broad
redshift probability distributions, due in part to insufficient
depth in photometry at shorter wavelengths.  Simulations indicate
a factor of two reduction in IRAC photometric error can help compensate,
leading to a similar reduction in photometric redshift error in this
redshift range. With the {\it Spitzer} Deep Wide-Field Survey
legacy program now underway (PI D.~Stern), such data will be available in the near future.  
The tighter redshift probability distributions
would improve the contrast of high redshift clusters over the field,
allowing them to meet our detection threshold. 

But the lack of massive $z > 1.5$ cluster
detections may simply reflect a real decline in the 
space density of such objects.  Models for the hierarchical
growth of structure predict only about 1 cluster with $z> 1.4$ 
above $10^{14}  M_{\odot}$ for every 5 clusters with $z > 1$.
Full exploration of this possibility will require more
careful simulation and assessment of our observational
selection effects.

\section{Summary\label{sec:summary}}

We have identified 335 galaxy cluster and group candidates from a 
$4.5 \mu$m selected sample of galaxies in the IRAC Shallow Survey.
Candidates were identified by searching for overdensities in photometric
redshift slices, 
and 106 clusters are at $z > 1$.   Roughly 10\% of these candidates
may be expected to arise by chance or from projection effects.
To date, 12 clusters have
been spectroscopically confirmed at $z > 1$, as have 61 of the 73 clusters
observed with AGES at $z < 0.5$.  
For the two $z > 1$ clusters with 20 or more spectroscopic
members, total cluster masses of several $10^{14} {\rm  M_\odot}$ are indicated, 
and the total mass estimated from the stellar luminosity yields comparable values.  
Color-magnitude diagrams in $I - [3.6]$ vs. [3.6]
for the $z > 1$ spectroscopically confirmed clusters reveal that 
a red sequence is generally present, even though clusters were not selected for this.  
The brightest probable member galaxy (at the spatial resolution of IRAC) 
in the spectroscopically confirmed $ z > 1$ 
clusters remains 1 -- 2 mag brighter than
the passively-evolving $L^*$ luminosity.
For the full cluster sample, the mean color of
brighter galaxies within each cluster is 
systematically redder than the mean color of all probable cluster member galaxies,
implying that the mass-metallicity relation is already in place at $z \sim 1.5$.  
The mean $I - [3.6]$ color of probable cluster members is well fit
by a simple model in which stars form in a 0.1 Gyr burst beginning
at $z_f = 3$. At $z > 1$ there is a tendency for mean cluster colors to
favor formation redshifts $z_f > 3$, although a few are consistent with
$z_f \sim 2$.  
This adds to the large body of evidence that galaxies in clusters 
were established at extremely early times.  

\acknowledgements 

We thank Mark Dickinson, Emily MacDonald, and Hyron Spinrad 
for generously making time available
on their scheduled nights for the DEIMOS observations reported here. 
Naoki Yasuda, Naohiro Takanashi, Yutaka Ihara, Kohki Konishi, and 
Hiroyuki Utsunomiya assisted with observations at the Subaru Telescope.
Roberto De Propris provided the integrated luminosity for Coma cluster galaxies.
Thoughtful comments from the anonymous referee improved the presentation
of this work.  
The IRAC Shallow Survey was executed using guaranteed observing
time contributed by G. Fazio, G. and M. Rieke, M. Werner, and E. Wright.  
This work is based in part on observations made with 
the {\it Spitzer Space Telescope}, which is operated by 
the Jet Propulsion Laboratory, California Institute of Technology 
under a contract with NASA. 
This work made use of images and data products provided by the 
NOAO Deep Wide-Field Survey, which is supported by the 
National Optical Astronomy Observatory (NOAO). 
NOAO is operated by AURA, Inc., 
under a cooperative agreement with the National Science Foundation.
Some of the data presented herein were obtained at the W.M. Keck Observatory, 
which is operated as a scientific partnership among the 
California Institute of Technology, the University of California 
and the National Aeronautics and Space Administration. 
The Observatory was made possible by the generous financial support 
of the W.M. Keck Foundation.  
Some of the data presented were collected at the Subaru Telescope, which is 
operated by the National Astronomical Observatory of Japan.
The work of SAS was performed under the auspices
of the U.S. Department of Energy, National Nuclear Security Administration,
by the University of California, Lawrence Livermore National Laboratory,
under contract No. W-7405-Eng-48.
The work of KD and JM was partially supported by the Director, Office of Science, 
Department of Energy, under grant DE-AC02-05CH11231.

\bibliographystyle{apj}

\begin{deluxetable}{lccc}
\tabletypesize{\normalsize}
\tablecaption{$z > 1$ Spectroscopic Cluster Members\label{SpecTable}}
\tablewidth{0pt}
\tablehead{
\multicolumn{1}{c}{ID\tablenotemark{a}} & 
\multicolumn{1}{c}{$[4.5]$\tablenotemark{b}}&
\multicolumn{1}{c}{phot--$z$} &
\multicolumn{1}{c}{spec--$z$}
}
\startdata
\multicolumn{4}{c}{ISCS\_J1434.1+3328 $\left<z_{\rm sp}\right> = 1.057$} \\
\hline
152.1 & 15.45  &  1.01  & 1.057 \\
152.2 & 15.99  &  0.91  & 1.055 \\
152.3 & 17.00  &  0.96  & 1.054 \\
152.4 & 17.47  &  1.04  & 1.065 \\
152.5 & 16.40  &  0.99  & 1.055 \\
152.6 & 16.76  &  0.96  & 1.055 \\
\hline
\multicolumn{4}{c}{ISCS\_J1429.2+3357 $\left<z_{\rm sp}\right> = 1.058$} \\
\hline
51.1  & 15.10  &  0.98  & 1.059 \\  
51.2  & 17.06  &  0.95  & 1.056 \\  
51.3  & 17.83  &  1.04  & 1.055 \\  
51.4  & 17.16  &  1.00  & 1.060 \\  
51.5  & 17.01  &  1.03  & 1.059 \\  
51.6  & 17.19  &  1.03  & 1.054 \\  
51.7  & 16.08  &  1.01  & 1.060 \\ 
\hline 
\multicolumn{4}{c}{ISCS\_J1433.1+3334 $\left<z_{\rm sp}\right> = 1.070$} \\
\hline
19.1  & 17.49  &  0.99  & 1.075 \\ 
19.2  & 16.67  &  0.99  & 1.075 \\ 
19.3  & 17.04  &  1.06  & 1.076 \\ 
19.4  & 16.00  &  1.06  & 1.066 \\ 
19.5  & 16.89  &  1.09  & 1.067 \\ 
19.6  & 16.47  &  0.94  & 1.065 \\ 
19.7  & 17.70  &  1.00  & 1.063 \\ 
19.8  & 17.15  &  1.10  & 1.079 \\ 
19.9  & 16.75  &  0.98  & 1.074 \\ 
19.10  & 16.30  &  1.07  & 1.066 \\ 
19.11  & 16.21  &  1.03  & 1.066 \\ 
19.12  & 16.66  &  1.03  & 1.075 \\ 
19.13  & 15.44  &  1.06  & 1.064 \\ 
19.14  & 17.03  &  1.28  & 1.076 \\ 
19.15  & 17.51  &  1.00  & 1.065 \\ 
19.16  & 17.01  &  1.27  & 1.069 \\ 
19.17  & 17.36  &  1.04  & 1.066 \\ 
19.18  & 16.27  &  0.86  & 1.063 \\ 
19.19  & 16.68  &  2.45  & 1.074 \\ 
19.20\tablenotemark{c}  &   -    &   -    & 1.067 \\  
\hline
\multicolumn{4}{c}{ISCS\_J1433.2+3324 $\left<z_{\rm sp}\right> = 1.096$} \\
\hline
123.1  & 15.32  &  1.07  & 1.094 \\  
123.2  & 16.45  &  1.09  & 1.093 \\  
123.3  & 16.68  &  1.10  & 1.094 \\  
123.4  & 16.99  &  1.11  & 1.107 \\  
123.5  & 16.64  &  0.93  & 1.094 \\  
123.6  & 17.35  &  0.99  & 1.091 \\  
\hline
\multicolumn{4}{c}{ISCS\_J1432.4+3332 $\left<z_{\rm sp}\right> = 1.112$} \\
\hline
17.1  & 16.94  &  2.74  & 1.1120\\  
17.2  & 17.35  &  1.03  & 1.1108\\  
17.3  & 16.48  &  1.07  & 1.111 \\  
17.4  & 15.97  &  1.05  & 1.115 \\  
17.5  & 16.31  &  1.04  & 1.111 \\  
17.6  & 15.85  &  1.09  & 1.110 \\  
17.7  & 16.04  &  1.35  & 1.121 \\  
17.8  & 17.57  &  1.20  & 1.116 \\  
17.9  & 16.87  &  1.00  & 1.11  \\  
17.10  & 17.31  &  1.04  & 1.1086\\  
17.11  & 17.01  &  1.14  & 1.098 \\  
17.12  & 17.43  &  1.12  & 1.105 \\  
17.13  & 17.51  &  0.98  & 1.109 \\  
17.14  & 17.32  &  0.91  & 1.112 \\  
17.15  & 16.59  &  1.07  & 1.104 \\  
17.16  & 16.92  &  1.19  & 1.119 \\  
17.17  & 17.42  &  1.09  & 1.115 \\  
17.18  & 17.33  &  1.11  & 1.115 \\  
17.19  & 17.29  &  1.06  & 1.118 \\  
17.20  & 16.90  &  1.18  & 1.107 \\  
17.21  & 16.38  &  1.26  & 1.115 \\  
17.22\tablenotemark{c}  &  -    &   -    & 1.114 \\  
17.23\tablenotemark{c}  &  -    &   -    & 1.110 \\ 
\hline 
\multicolumn{4}{c}{ISCS\_J1426.1+3403 $\left<z_{\rm sp}\right> = 1.135$} \\
\hline
34.1  & 16.35  &  1.03  & 1.1439 \\ 
34.2  & 15.75  &  1.04  & 1.1301 \\ 
34.3  & 16.01  &  1.11  & 1.1271 \\ 
34.4  & 15.69  &  1.00  & 1.1328 \\ 
34.5  & 16.12  &  1.18  & 1.134 \\  
34.6  & 16.90  &  0.91  & 1.13 \\  
34.7  & 17.24  &  0.89  & 1.144 \\ 
\hline
\multicolumn{4}{c}{ISCS\_J1426.5+3339 $\left<z_{\rm sp}\right> = 1.161$} \\
\hline
14.1  & 16.87  &  1.10  & 1.16   \\ 
14.2  & 15.96  &  1.05  & 1.157  \\ 
14.3  & 16.44  &  1.19  & 1.1631 \\ 
14.4  & 16.64  &  1.17  & 1.1634 \\ 
14.5  & 16.88  &  1.04  & 1.1637 \\ 
\hline
\multicolumn{4}{c}{ISCS\_J1434.5+3427 $\left<z_{\rm sp}\right> = 1.243$ \tablenotemark{d}}  \\
\hline
342.1  & 16.76  &  1.21  & 1.240 \\  
342.2  & 17.71  &  1.21  & 1.251 \\  
342.3\tablenotemark{c}  &   -    &   -    & 1.256 \\  
\hline
\multicolumn{4}{c}{ISCS\_J1429.3+3437 $\left<z_{\rm sp}\right> = 1.258$}  \\
\hline
30.1  & 16.65  &  1.18  & 1.245 \\
30.2  & 16.39  &  1.13  & 1.26  \\
30.3  & 17.24  &  1.15  & 1.2576\\
30.4  & 17.37  &  1.07  & 1.263 \\
30.5  & 17.41  &  1.29  & 1.2611\\
30.6  & 16.86  &  1.08  & 1.2583\\
30.7  & 16.97  &  1.31  & 1.2582\\
30.8  & 15.19  &  0.97  & 1.2632\\
30.9\tablenotemark{c}  &   -    &   -    & 1.2546\\
\hline
\multicolumn{4}{c}{ISCS\_J1432.6+3436 $\left<z_{\rm sp}\right> = 1.347$} \\
\hline
29.1  & 17.63  &  1.26  & 1.3559 \\  
29.2  & 17.61  &  1.25  & 1.35 \\  
29.3  & 16.39  &  1.32  & 1.35 \\  
29.4  & 16.06  &  1.21  & 1.3320 \\
29.5  & 17.55  &  1.40  & 1.347 \\ 
29.6  & 16.77  &  1.29  & 1.347 \\ 
29.7  & 16.86  &  1.30  & 1.34 \\  
29.8  & 16.14  &  1.26  & 1.353 \\
\hline 
\multicolumn{4}{c}{ISCS\_J1434.7+3519 $\left<z_{\rm sp}\right> = 1.373$}  \\
\hline
25.1  & 17.52  &  3.33  & 1.37  \\  
25.2  & 16.98  &  1.53  & 1.372 \\  
25.3  & 16.28  &  1.34  & 1.37  \\  
25.4  & 15.68  &  1.33  & 1.374 \\  
25.5\tablenotemark{c}  &   -    &   -    & 1.380 \\ 
\hline 
\multicolumn{4}{c}{ISCS\_J1438.1+3414 $\left<z_{\rm sp}\right> = 1.413$ \tablenotemark{e}} \\
\hline
22.1  & 15.86  &  1.25  & 1.411 \\  
22.2  & 15.79  &  1.38  & 1.418 \\  
22.3  & 16.91  &  1.38  & 1.412 \\  
22.4  & 16.76  &  1.35  & 1.414 \\  
22.5\tablenotemark{c}  &   -    &   -    & 1.412 \\  
\enddata
\tablenotetext{a}{Coordinates will appear in published version.}
\tablenotetext{b}{Vega magnitude at 4.5\mum; 0 mag = 179.5 Jy.}
\tablenotetext{c}{Serendipitous spectroscopic source not included in 5$\sigma$ 4.5$\mu$m catalog.} 
\tablenotetext{d}{Additional spectroscopic members for ISCS J1434.5+3427 published in \citet{Brodwin2006}}
\tablenotetext{e}{Additional spectroscopic members for ISCS J1438.1+3414 published in \citet{Stanford2005}}
\end{deluxetable}

\end{document}